\documentclass[pre,twocolumn,10pt,aps,superscriptaddress]{revtex4-1}
\usepackage[T1]{fontenc}
\usepackage[utf8]{inputenc}
\usepackage[english]{babel}
\usepackage{amsmath}
\usepackage{amssymb}
\usepackage{amsthm}
\usepackage{dsfont}
\usepackage{amsfonts}
\usepackage{xcolor}
\usepackage{graphicx}
\usepackage{array}
\usepackage{ragged2e}
\usepackage{url}
\usepackage{import}
\usepackage{soul}
\usepackage{url}
\usepackage{hyperref}
\usepackage{ulem}
\usepackage{cancel}
\usepackage{extarrows}
\usepackage{enumitem}

\newcommand{\bra}[1]{\langle #1|}
\newcommand{\ket}[1]{|#1\rangle}

\newcommand{\avin}[3]{\langle #1|#2|#3 \rangle}

\begin{document}

\title{Self-Consistent Density-Functional Embedding: a Novel Approach for Density-Functional Approximations}

\author{Uliana Mordovina}
\email{uliana.mordovina@mpsd.mpg.de \\ U.~M. and T.~E.~R. contributed equally}
\affiliation{Max Planck Institute for the Structure and Dynamics of Matter, Center for Free Electron Laser Science, 22761 Hamburg, Germany}

\author{Teresa E.~Reinhard}
\email{uliana.mordovina@mpsd.mpg.de \\ U.~M. and T.~E.~R. contributed equally}
\affiliation{Max Planck Institute for the Structure and Dynamics of Matter, Center for Free Electron Laser Science, 22761 Hamburg, Germany}

\author{Iris Theophilou}
\affiliation{Max Planck Institute for the Structure and Dynamics of Matter, Center for Free Electron Laser Science, 22761 Hamburg, Germany}

\author{Heiko Appel}
\email{heiko.appel@mpsd.mpg.de}
\affiliation{Max Planck Institute for the Structure and Dynamics of Matter, Center for Free Electron Laser Science, 22761 Hamburg, Germany}

\author{Angel Rubio}
\email{angel.rubio@mpsd.mpg.de}
\affiliation{Max Planck Institute for the Structure and Dynamics of Matter, Center for Free Electron Laser Science, 22761 Hamburg, Germany}
\affiliation{Center for Computational Quantum Physics (CCQ), Flatiron Institute, 162 Fifth Avenue, New York NY 10010, USA}

\date{\today}

\begin{abstract}
In the present work, we introduce a Self-Consistent Density-Functional Embedding technique, which leaves the realm of standard energy-functional approaches in Density Functional Theory and targets directly the density-to-potential mapping that lies at its heart. Inspired by the Density Matrix Embedding Theory, we project the full system onto a set of small interacting fragments that can be solved accurately. Based on the rigorous relation of density and potential in Density Functional Theory, we then invert the fragment densities to local potentials. Combining these results in a continuous manner provides an update for the Kohn-Sham potential of the full system, which is then used to update the projection. The scheme proposed here converges to an accurate approximation for the density and the Kohn-Sham potential of the full system. Convergence to exact results can be achieved by increasing the fragment size. We find, however, that already for small embedded fragments accurate results are obtained. We benchmark our approach for molecular bond stretching in one and two dimensions and demonstrate that it reproduces the known steps and peaks that are present in the exact exchange-correlation potential with remarkable accuracy.
\end{abstract}

\maketitle

\section{Introduction}
Over the past decades, Density Functional Theory (DFT) has become a well established and
successful method able to accurately describe molecular and condensed matter systems.
One reason for its success can be attributed to its computational efficiency as all physical observables of interest are functionals
of the ground-state density $n({\bf r})$\cite{hohenberg_inhomogeneous_1964}. 
The most popular technique to find the density of the system accurately is the Kohn-Sham (KS) DFT, where the density of the full interacting system
is computed via an auxiliary non-interacting system \cite{kohn_self-consistent_1965}. 
All interactions and correlations of the interacting system are mimicked by the so-called exchange-correlation (xc) potential which is usually determined as the derivative of the xc energy functional $E_{\rm xc}[n]$ that is unknown and has to be approximated in practice
\cite{kohn_self-consistent_1965,perdew_generalized_1996,becke_density-functional_1988,becke_densityfunctional_1993,lee_development_1988}.
A remaining challenge is to find functional approximations describing other wanted observables $O[n]$.

Another issue with DFT is that, although significant progress in functional development over the years has been achieved, 
approximate DFT functionals usually still struggle to describe
systems with strongly correlated electrons \cite{cohen2011challenges}. The dissociation limit of the
$H_2$ molecule is a good example for a simple system that is not easy to
describe with commonly used approximate DFT functionals. Those functionals that are optimized to be able to mimic the dissociation of $H_2$\cite{heselmann_correct_2011,vuckovic_hydrogen_2015,zhao2008,zhao2011,leininger1997,toulouse2005} do not perform equally good on other problems\cite{cohen2008insights}. 

There are alternative methods that are able to describe strongly correlated electrons accurately. One big group are wave function methods, such as full configuration interaction
(FCI) methods \cite{szabo_modern_1996} and density-matrix-renormalization group (DMRG)
\cite{schollwoeck_density-matrix_2005}.  These methods, although becoming more
and more efficient, still have high computational cost and thus are only able to describe relatively small systems.

A pathway to use accurate methods on a larger scale is provided by embedding theories. The general idea behind embedding consists of dividing a system into one or more
fragments  of interest and an environment, which is then considered only indirectly. With this partition the need of performing an expensive calculation on the full system is circumvented.
An established group of embedding theories are various density-functional embedding methods \cite{cortona1991self,wesolowski1993frozen,govind1999electronic,fulde2012electron,manby2012simple,jacob2014subsystem}
that have been successfully applied to a large range of complex systems, such as molecule adsorption on metallic surfaces\cite{kluner2001prediction}, proton transfer reactions in solution\cite{xiang2008quantifying} and photophysical properties of natural light-harvesting complexes\cite{neugebauer2008photophysical} - to name a few. They provide ways of calculating a system which is weakly bounded to an environment
by representing the environment by an external field. Opposed to that, embedding methods such as dynamical-mean-field theory (DMFT) \cite{georges_dynamical_1996,kotliar_electronic_2006,held_electronic_2007},
density-matrix-embedding theory (DMET)\cite{knizia_density_2012,knizia_density_2013,wouters_practical_2016}, and density-embedding theory (DET)\cite{bulik_density_2014,bulik_electron_2014} 
consider correlations between system and environment more explicitly and, thus, are successful in describing systems with
strongly correlated electrons. This is achieved though mapping the full system onto a fragment that is embedded into a, in some cases correlated, bath. In the latter two
methods, only the fragment is described accurately while the rest of
the system is described with a lower level calculation. Here, the challenge is
the connection between the high-level and the low-level calculation
\cite{bulik_electron_2014,welborn_bootstrap_2016,bulik_density_2014,wouters_practical_2016,knizia_density_2012,booth_spectral_2015,wu_p-DMET_2019}.

All mentioned embedding methods are tailored to describe the behavior of the fragments accurately.
Opposed to that, we use in the present work the embedding idea to improve our large scale description of the full system by including
insights from small fragments. To this end we introduce a feedback algorithm, which combines DMET with
density inversions based on the one-to-one correspondence of density and potential in exact DFT. This results in a self-consistent density-functional embedding (SDE) technique,
which allows to explicitly construct approximations to the xc potential with increasing accuracy. Here, no optimized-effective potential (OEP)\cite{sharp1953variational,talman1976optimized} procedure needs to be employed
since it is not the energy that is approximated, but directly the local xc potential. In our context we are not using an explicit expression of the xc potential in terms of the density but rather employ a direct numerical construction.
We use the embedding to find numerically local approximations to the density-potential mappings that give direct access to the xc potential.
Once the optimal KS system is obtained, we gain information about observables from those interacting fragment wave functions, which serve as an approximation to the full interacting wave function. To put it differently, we approximate the involved potential-density maps of standard DFT as combination of local maps. In the limit that the fragment describes the full system, we find also the exact KS potential.

The paper is organized as follows. In section \ref{sec:theory}, we introduce
the proposed SDE method step by step.  In section \ref{section:H2}, we present
the Hamiltonian for two electrons in a heteroatomic model system in one and two
dimensions, which we use to benchmark our approach. The results for the energy, the density
and the KS potential of the introduced systems are shown in section
\ref{sec:results}. Finally, a summary of the SDE method and an outlook towards more general applications is given in section \ref{sec:sum}.

\section{Theory}\label{sec:theory}
\subsection{Density Functional Theory} \label{sec:dft}

In this section we introduce key aspects of DFT and issues of standard DFT approximations that we wish to address with our approach. Based on the Hohenberg-Kohn (HK) theorem\cite{hohenberg_inhomogeneous_1964}, in KS DFT\cite{kohn_self-consistent_1965} the ground-state density $n(\boldsymbol{r})$ of a target interacting many-body system is obtained through a set auxiliary one-body (KS) equations with an effective local (KS) potential $v_{\rm KS}[n](\boldsymbol{r})$ (atomic units are used throughout the paper)
\begin{align}
	\left( - \frac{\boldsymbol{\nabla}^2}{2}  +  v_\mathrm{KS}(\boldsymbol{r}) \right) \varphi_j(\boldsymbol{r}) & = \varepsilon_j \varphi_j(\boldsymbol{r}), \label{eq:ks_eq}\\ 
\end{align}
The difference between the KS potential and the external potential $v_{\rm ext}(\boldsymbol{r})$ of the interacting system is the so-called Hartree-exchange-correlation (Hxc) potential $v_{\rm Hxc}[n](\boldsymbol{r})$ that accounts for all the interactions and the kinetic correlations of the interacting system. This potential is usually obtained by approximating the corresponding Hxc energy functional $E_{\rm Hxc}[n]$ and then taking the functional derivative of the latter with respect to the density.

Although highly successful, there are several issues with this approach. From a formal perspective, it has be shown that the exact functionals as defined by Lieb are not functionally differentiable \cite{Lammert2007}. So, to provide the main ingredient, regularizations need to be done \cite{KSconvergence2019}. Further, it is very hard to systematically increase the accuracy of known approximate functionals \cite{Medvedev49}. And, even if we had an accurate approximate functional,
it would usually be given in terms of KS orbitals and a numerically demanding OEP procedure\cite{sharp1953variational,talman1976optimized} would be needed to obtain the KS potential. 
Furthermore, there is the often overlooked but important issue of how to construct other observables from the KS Slater determinant as any observable that cannot be expressed directly in terms of the density needs to be approximated in terms of the latter.\par

Here, we avoid these issues by following a different path which involves no explicit approximate expression for $E_{\rm Hxc}[n]$ or $v_{\rm Hxc}[n]$. Instead, we first introduce a formal approach that employs density-potential mappings of DFT directly (see e.g. \cite{ruggenthaler2015existence}) and then make this approach practical by applying approximations to it. Following the HK theorem, for a given density $n^{(i)}$ there is a interacting system with the external potential $v[n^{(i)}]$ that produces this density. And exactly the same density can be reproduced by non-interacting system with the potential $v_s[n^{(i)}]$. Hence, an interacting density $n^{(i)}(\boldsymbol{r})$ can be uniquely inverted to both an interacting potential $v[n^{(i)}](\boldsymbol{r})$ and a non-interacting potential $v_s[n^{(i)}](\boldsymbol{r})$ . The Hxc potential is then defined by the difference of those two potentials
\begin{equation}
    v_\mathrm{KS}[v_\mathrm{ext}, n^{(i)}](\boldsymbol{r})  =v_\mathrm{ext}(\boldsymbol{r}) + \underbrace{v_{\rm S}[n^{(i)}](\boldsymbol{r})-v[n^{(i)}](\boldsymbol{r})}_{v_{\rm Hxc}[n^{(i)}](\mathbf{r})}.
    \label{eq:fixed-point}
\end{equation}
Solving the single-particle eigenvalue equations eq~(\ref{eq:ks_eq}) for $v_\mathrm{KS}[v_\mathrm{ext}, n^{(i)}]$ we obtain the updated density $n^{(i+1)}$. Starting with some initial density $n^{(0)}$, this scheme converges at the true ground state density $n$ that is produced by the external potential $v_{\rm ext} = v[n]$ and we have also found the non-interacting potential $v_{\rm KS}[n] = v_{\rm S}[n]$ to reproduce this density.

Note that the fixed-point iteration scheme above does not need any explicit expression of an energy functional. However, it is obvious that the scheme itself is not practical at all. In order to avoid solving the exact Schr\"odinger equation (SE) for one interacting system with $v_{\rm ext}$ we ended up performing inversions not only to obtain the non-interacting $v_{\rm S}[n^{(i)}]$, which in principle is feasible\cite{kananenka_efficient_2013,Jensen2017,Nielsen2018}, but also to obtain the interacting $v[n^{(i)}]$, which would involve solving the SE multiple times at each step and, hence, increase the numerical complexity of the problem instead of decreasing it.

The method we present in this paper targets directly at approximating the fixed-point iteration scheme in a way that no inversion for $v[n]$ is necessary. Within our approach the connection between $v[n]$ and $n$ is given by a projection (that we introduce in subsection \ref{sec:how_to}) and the exact SE is solved in smaller subsystems.

\subsection{Self-Consistent Density-Functional Embedding Method}

The fundamental idea of the SDE approach is to replace the mapping between the global KS potential and the corresponding density by dividing the system into a set of fragments $\left\{i\right\}$ and mapping those onto a set of auxiliary interacting systems with a corresponding set of external potentials $\left\{v^i\right\}$, interacting wave-functions $\left\{\ket{\Psi}^i\right\}$ and densities $\left\{n^i\right\}$. Here, no interacting inversion is needed and we also get an approximated mapping between the KS Slater determinant $\ket{\Phi_0}$ and the ground-state wave function of the system $\ket{\Psi_0}$.


\begin{figure*}[t]
	\includegraphics[width=\textwidth]{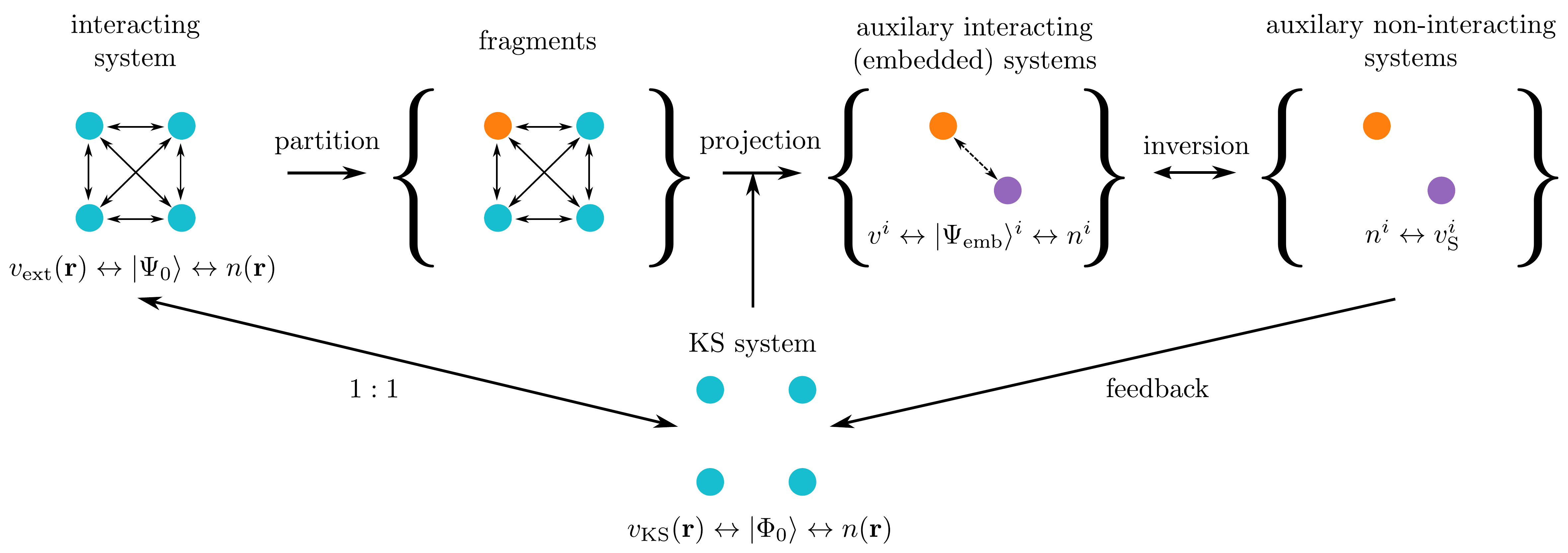}
        \caption{General SDE idea: properties of an interacting electronic system with an external potential $v_{\rm ext}$ and a ground-state wave function $\ket{\Psi_0}$ are fully determined by its electronic density $n(\boldsymbol{r})$, that can be uniquely reproduced by a non-interacting system (KS system).
        The interacting system is divided into fragments. For each fragment (orange) the system is projected onto a smaller auxiliary interacting (embedded) system.
        The embedded system consists of the fragment, which remains unchanged by the projection and the part of the system that includes interaction and correlation with the fragment (depicted in violet). Each of the embedded systems is then solved on a wave-function level, yielding an accurate density which then can be uniquely mapped onto an auxiliary non-interacting system with the same density. 
        These accurate local potentials are then used to improve the global KS description of the full system. The whole process is repeated self-consistently until convergence of the global KS potential is reached.}
	\label{pic:general_emb}
\end{figure*}

The SDE method is depicted schematically in Fig.~\ref{pic:general_emb}. It consists of the following parts, to each of which we assign a distinct subsection:

\begin{enumerate}
 \item The full system is described in terms of its ground-state density $n(\boldsymbol{r})$ by means of KS DFT, as we have discussed in subsection \ref{sec:dft}.
 \item The system is divided into fragments. Our proposed partition differs significantly from partition DFT\cite{elliott2010partition,nafziger2014density} or DMET\cite{wouters_practical_2016} and we will introduce our 'continuous partition'  in subsection \ref{sec:patching}.
 \item For each fragment, the full system is projected onto an embedded system, where the fragment is embedded into an effective bath. In this paper, the choice for the projector is inspired by the DMET approach, which we explain in detail in subsection \ref{sec:how_to}. 
 
 \item For each fragment, an accurate calculation is performed with a wave-function method.
 The fragment wave functions are then used to calculate accurate fragment densities. These wave functions also serve as a local approximation to the mapping between the KS Slater determinant and the ground-state wave function $\ket{\Phi_0[n]} \rightarrow \ket{\Psi_0[n]}$, from which we can directly calculate correlated observables via $ O[n]= \avin{\Psi[n]}{\hat{O}}{\Psi[n]} $. We explain, how this calculation is performed in practice in section \ref{sec:fragment}.
 \item Finally, for each fragment $i$ an auxiliary non-interacting system is found that reproduces the density $n_i$ and the set of obtained potentials $\left\{v_S[n^i]\right\}$ is then used to update the global KS potential. How this is done in practice is explained in subsection \ref{sec:scf}. The SDE scheme is applied self-consistently and the algorithm is also explained in subsection \ref{sec:scf}.
\end{enumerate}

As we divide our system into fragments in real space, we will, for the sake of convenience, consider only systems that are discretized on a real space grid throughout the paper.

\subsection{Continuous partition}\label{sec:patching}
We continue by considering the problem of dividing the full problem into fragments.
Generally, the fragments have to cover the full system and should be selected small enough to be calculated with required accuracy.

In embedding approaches like subsystem DFT \cite{jacob2014subsystem} and also in the framework of partition DFT \cite{elliott2010partition},
the system is divided into non-overlapping fragments, which are weakly bounded to one another. In other words, the partition is dictated by density distribution and correlations within
the system and cannot be chosen arbitrarily. Therefore, those approaches are not applicable when connections along fragments become important.

\begin{figure}[t]
	\includegraphics[width=.4\textwidth]{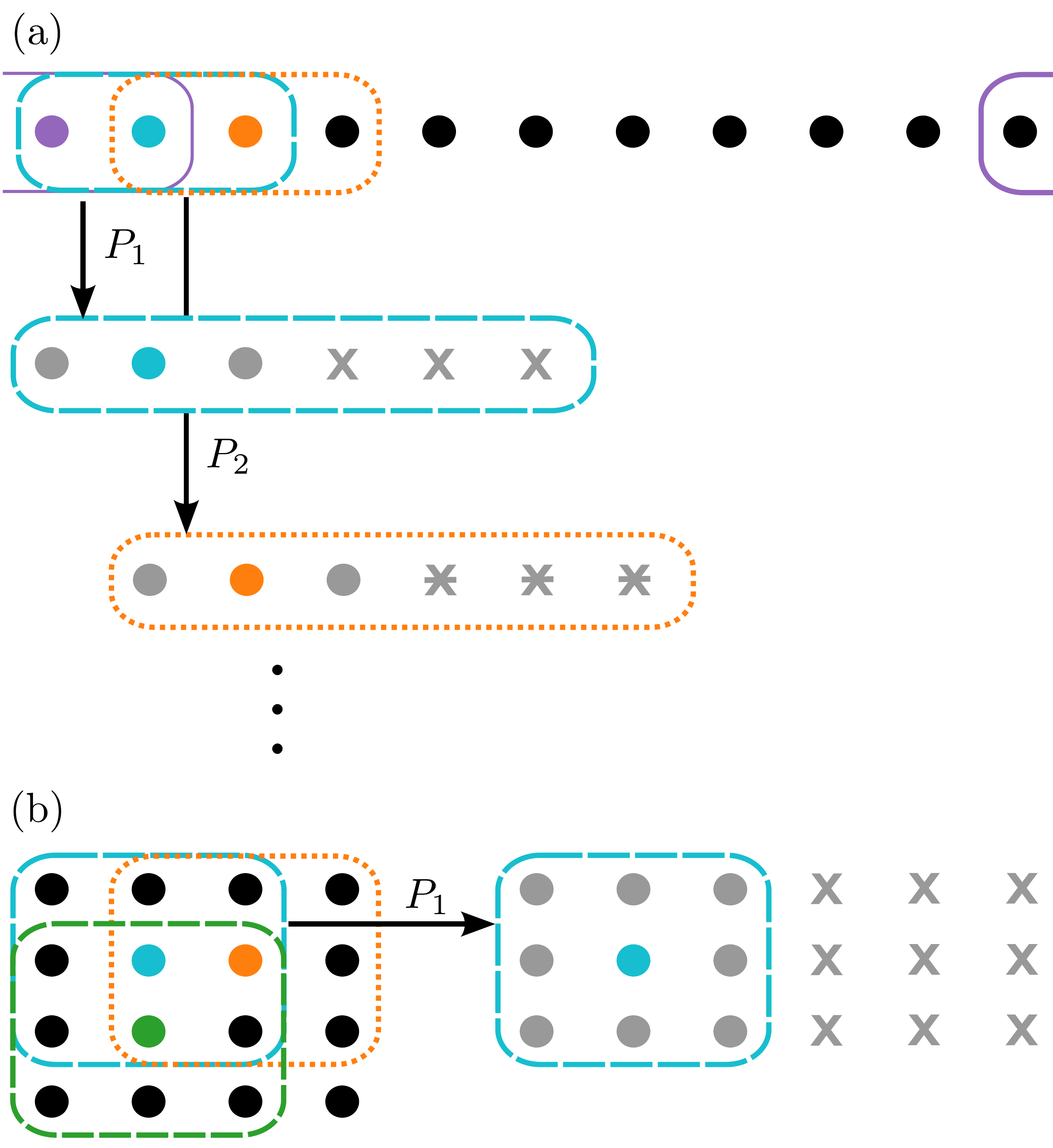}
        \caption{Visualization of the partition procedure: In order to obtain a
        continuous density, we sweep through the system by just going one site
        forward for each fragment calculation. Then, only the physical properties of the
        centering site are taken into account when considering local observables. The upper
        image (a) shows the partition in 1D, whereas the lower image (b)
        illustrates the partition in 2D. Projections $P_i$ onto embedded systems as well as effective bases depicted by different kinds of crosses are explained in section \ref{sec:how_to}.
        This partition procedure can be extended to 3D in a straight-forward manner (not shown).}
	\label{pic:patching}
\end{figure}

In DMET \cite{knizia_density_2012,knizia_density_2013,wouters_practical_2016} the system is also divided into non-overlapping fragments. The partition itself can be chosen arbitrarily, as particle transfer between fragment and the rest of the system is possible within this approach. The size of the fragments is dictated mostly by the correlation length in the system \cite{knizia_density_2012}. Hence, the
amount of correlation, which is captured with the DMET method is constrained by the size of the fragment. Thus, by increasing the fragment size, a convergence towards the exact solution is feasible, which makes the method systematically improvable. Dividing the system into non-overlapping fragments, however, causes artificial discontinuities in local observables such as density\cite{welborn_bootstrap_2016}\footnote{We show a numerical example in section \ref{sec:KS}}, which sometimes also leads to convergence problems\cite{wu_p-DMET_2019}. This is the reason why DMET, in general, cannot be applied to heterogeneous systems self-consistently\cite{wu_p-DMET_2019}. For such systems a simple single-shot embedding is usually performed\cite{wouters_practical_2016}, which still provides very good results for the energies, which is after all the target of the DMET method.

In SDE, we employ the same type of projection as in DMET (see section \ref{sec:how_to}) but, since we are particularly targeting the density, we further introduce a partition that guarantees that all fragments connect smoothly to one another. Specifically, we define a continuous partition, where the system is covered by overlapping fragments as is depicted in Figure \ref{pic:patching}. In practice, we sweep through the system by just going one grid point further for each fragment calculation.
When computing local observables such as the density, we only take into account the grid point in the middle of each fragment. Hence, our partition is constructed such
that the local observables are continuous on the real-space grid. The accuracy can be improved by selecting the grid spacing appropriately. In practice, this has to be balanced with the computational cost as for any real-space implementation.

\subsection{Projection onto the embedded system}\label{sec:how_to}

Having decided on how to divide the system into fragments, we now treat each fragment separately and find an effective description for the corresponding embedded system (see Figure \ref{pic:general_emb}).
We want the embedded system to be such that it describes the physics on the fragment as accurately as possible.
As depicted in Figure \ref{pic:general_emb}, we have to project the full system onto an embedded system for each fragment.

Out of a manifold of possible projections \cite{georges_dynamical_1996,knizia_density_2012,manby2012simple} we adopt here the
the projection used in DMET \cite{knizia_density_2012,knizia_density_2013,wouters_practical_2016}
as it provides an efficient way of including static correlations between fragment and the rest of the system, which we call bath from now on. 

The DMET method can be understood as a complete active space (CAS) calculation under the assumption that the fragment basis functions are always in the active space. What then remains to be found are the orbitals that build up the remaining part of the active space, which we here call the correlated bath. It is constructed such that it has the same number of orbitals as the fragment $N_{\rm frag}$.



\begin{figure}[t]
	\includegraphics[width=.4\textwidth]{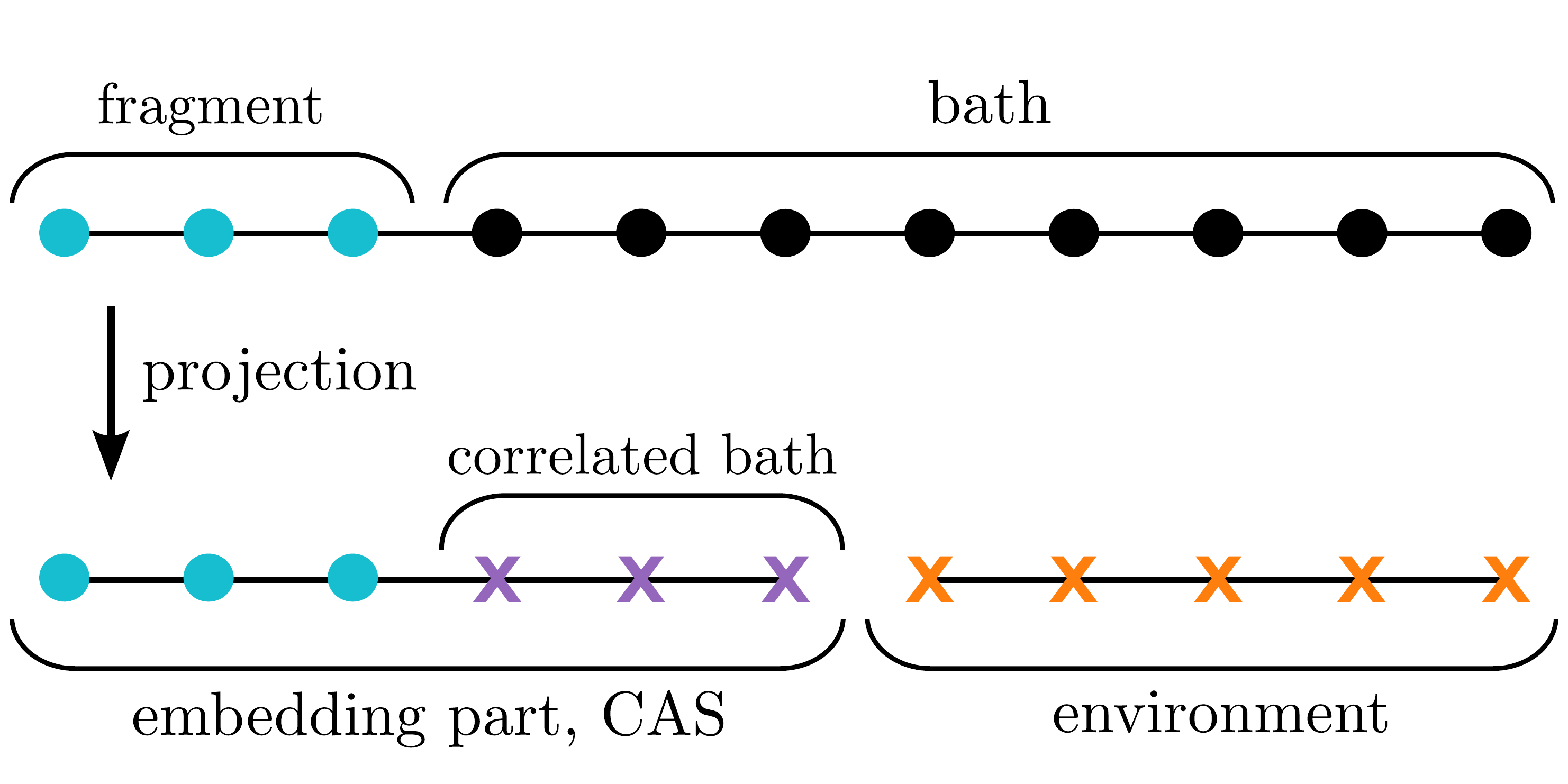}
        \caption{Visualization of the decomposition of the system into
        fragment and bath and the projection onto embedding (CAS) and environment
        part. The dots depict the sites, which correspond to our chosen initial basis set and the crosses the orbitals after projecting. In order to 
        describe the physics of the fragment only the embedding part is considered.}
	\label{pic:change_of_basis}
\end{figure}

Since the construction of the DMET projection has already been introduced in the literature multiple times\cite{knizia_density_2012,knizia_density_2013,wouters_practical_2016}, we leave the step-by-step instructions on how we do it in practice to the appendix~\ref{sec:appendix} and we give a visualization of the projection in Figure \ref{pic:change_of_basis}. By solving a mean-field Hamiltonian for the full system, we obtain a new smaller set of orbitals in which we then express the interacting Hamiltonian $\hat{H}$ of the full system to obtain the Hamiltonian $\hat{H}_{\rm emb}$ for the embedded system. 


Besides some technical subtleties that are discussed in the appendix~\ref{sec:appendix}, the main difference between the projector in SDE and DMET lies in the choice of the underlying non-interacting system. In DMET, any convenient non-interacting system can be chosen as its purpose lies only in approximating the projection. It is neither unique nor does it share actual observables with the target system. In SDE, on the other hand, the target non-interacting system is unique and well-defined. It is the KS system that reproduces the interacting density of the full system. 

\subsection{Fragment calculation}\label{sec:fragment}
For each fragment $i$ we obtain the embedding Hamiltonian $\hat{H}^i_{\rm emb}$ as described in section \ref{sec:how_to} and then diagonalize it to obtain the
embedding wave function $\ket{\Psi_{\rm emb}}^i$ and the corresponding density $n^i$ of this embedded system.

In the present work, we use exact diagonalization (ED) to solve for the ground-state
wave function of the embedded system. We emphasize that also
other solvers, such as DMRG \cite{schollwoeck_density-matrix_2005,hubig2017symmetry,chan2004algorithm},
coupled cluster \cite{bartlett2007coupled,sun2018pyscf,werner2012molpro}, selective CI approaches \cite{sharma2017semistochastic}, or Monte-Carlo
methods \cite{booth2009fermion,haule2007quantum} can be used for the fragment calculation.

The correlated embedding wave functions can then be used to calculate the energy of the full system $E_0$ or any other correlated observable.
As described in reference \cite{knizia_density_2013}, the energy of the full system $E_0$ can be approximated as a sum of fragment energies, which are calculated
by taking a partial trace of the corresponding embedding density matrix
$\hat{\rho}^i_{\rm emb} =\ket{\Psi_{\rm emb}}^i\bra{\Psi_{\rm emb}}^i$. In the SDE approach for each fragment $i$ only one site $\alpha_i$ is considered for obtaining properties
of the full system (see section \ref{sec:patching}). Therefore, to calculate observables, we redefine the fragment as only the local site of interest $\alpha_i$
and treat the remaining local sites together with the correlated bath orbitals as the rest of the embedded system.
Hence, we adopt the formula from reference \cite{knizia_density_2013} to
\begin{align}
E = \langle\hat{H}\rangle &\approx \sum\limits_i^N E_{\alpha_i} = \sum\limits_i^N \mathrm{Tr}_{\mathrm{CAS}-\alpha_i}\left(\hat{\rho}^i_{\rm emb} \hat{H}^i_{\rm emb} \right),
\label{eq:observable}
\end{align}
where $N$ denotes the number of grid points. Here, we have approximated the full wave function by a set of fragment wave-functions. The correlation length that can be captured within this approximation, is limited by the fragment size.

The formula above can be applied to any other observable. Thus, we circumvent the usual problem in DFT of finding explicit functional dependence $O[n]$ between an observable of interest $O$ and the density $n$
by simply using the embedding wave functions instead of the density.

Before moving on to improving the KS description of the full system, we have to add an additional constrain to the fragment calculations.
As in DMET or partition DFT, we have to make sure that, when patching the system back together, 
we retain the correct particle number $\mathcal{N}$ in the full system
\begin{equation}
 \langle\hat{\mathcal{N}}\rangle - \mathcal{N} \stackrel{!}{=} 0.
 \label{eq:part_num_constrain}
\end{equation}
Following reference \cite{wouters_practical_2016}, we achieve this by adding and self-consistently
optimizing a chemical potential $\mu$ to the embedding Hamiltonian of each fragment
\begin{equation}
 \hat{H}^i_\mathrm{emb} \rightarrow \hat{H}^i_\mathrm{emb} + \mu \sum_{\alpha \in N_\mathrm{frag}} \hat{n}_{\alpha},
 \label{eq:mu}
\end{equation}
where $\hat{n}_{\alpha}$ denotes the density operator on site $\alpha$ and the index $\alpha$ runs over all fragment sites.
The constant $\mu$ in eq~(\ref{eq:mu}) is added only to the fragment part of the embedding Hamiltonian in order
to achieve a correct particle distribution between fragment and environment.
In other words, the chemical potential is a Lagrange multiplier, which assures that
the constraint in eq~(\ref{eq:part_num_constrain}) is fulfilled.

\subsection{Self-consistency}\label{sec:scf}

So far, we have discussed how, starting from an initial guess for the KS potential (we usually start with $v_{\rm KS} = v_{\rm ext}$), we project the full system onto a set of interacting embedded systems with $\left\{H^i_{\rm emb} \leftrightarrow \ket{\Psi_{\rm emb}}^i \leftrightarrow n^i_{\rm emb}\right\}$. We now want to use this set of quantities to update the KS potential of the full system.

For each fragment $i$ the Hamiltonian contains a one-body part $\hat{h}^{i}_{\rm emb}$ and a two-body part $\hat{W}^{i}_{\rm emb}$
\begin{equation}
        \hat{H}^{i}_{\rm emb} = \hat{h}^{i}_{\rm emb} +\hat{W}^{i}_{\rm emb},
\end{equation}

Following the KS construction, the corresponding density $n^i_{\rm emb}$ can be reproduced by an auxiliary non-interacting system with
\begin{equation}
        \hat{H}^{i}_{\rm emb, \, MF} = \hat{h}^{i}_{\rm emb} +\hat{v}^{i}_{\rm emb,\, Hxc}[n^i_{\rm emb}]
 \end{equation}
where the correlations are mimicked by the Hxc potential $\hat{v}^{i}_{\rm Hxc,\,emb}$, that is defined as the difference of one-body terms of the interacting and the non-interacting systems. In practice, this potential is obtained either by analytical\cite{helbig_exact_2009} or numerical inversion\cite{kananenka_efficient_2013,Jensen2017,Nielsen2018}, or by a robust minimization routine as usually employed in DMET \cite{wouters_practical_2016}. The specific inversion scheme that is used to compute the results presented later in this paper will be introduced in section~\ref{section:H2} together with the model Hamiltonians we use for our results section.

We then approximate the Hxc potential of the full system $v_{\rm Hxc}$ on each site $\alpha_i$ by the corresponding value of $\hat{v}^{i}_{\rm Hxc,\,emb}$ on the same site.
\begin{equation}
    v_{\rm Hxc}(\alpha_i) = \hat{v}^{i}_{\rm Hxc,\,emb}(\alpha_i)
\end{equation}

The KS potential is then updated according to eq~(\ref{eq:fixed-point}) as
\begin{equation}
    \hat{v}_\mathrm{KS}(\alpha_i) = \hat{v}_{\rm ext}(\alpha_i) + \hat{v}^i_{\rm Hxc}(\alpha_i).
\end{equation}

\begin{figure}[t]
	\includegraphics[width=.5\textwidth]{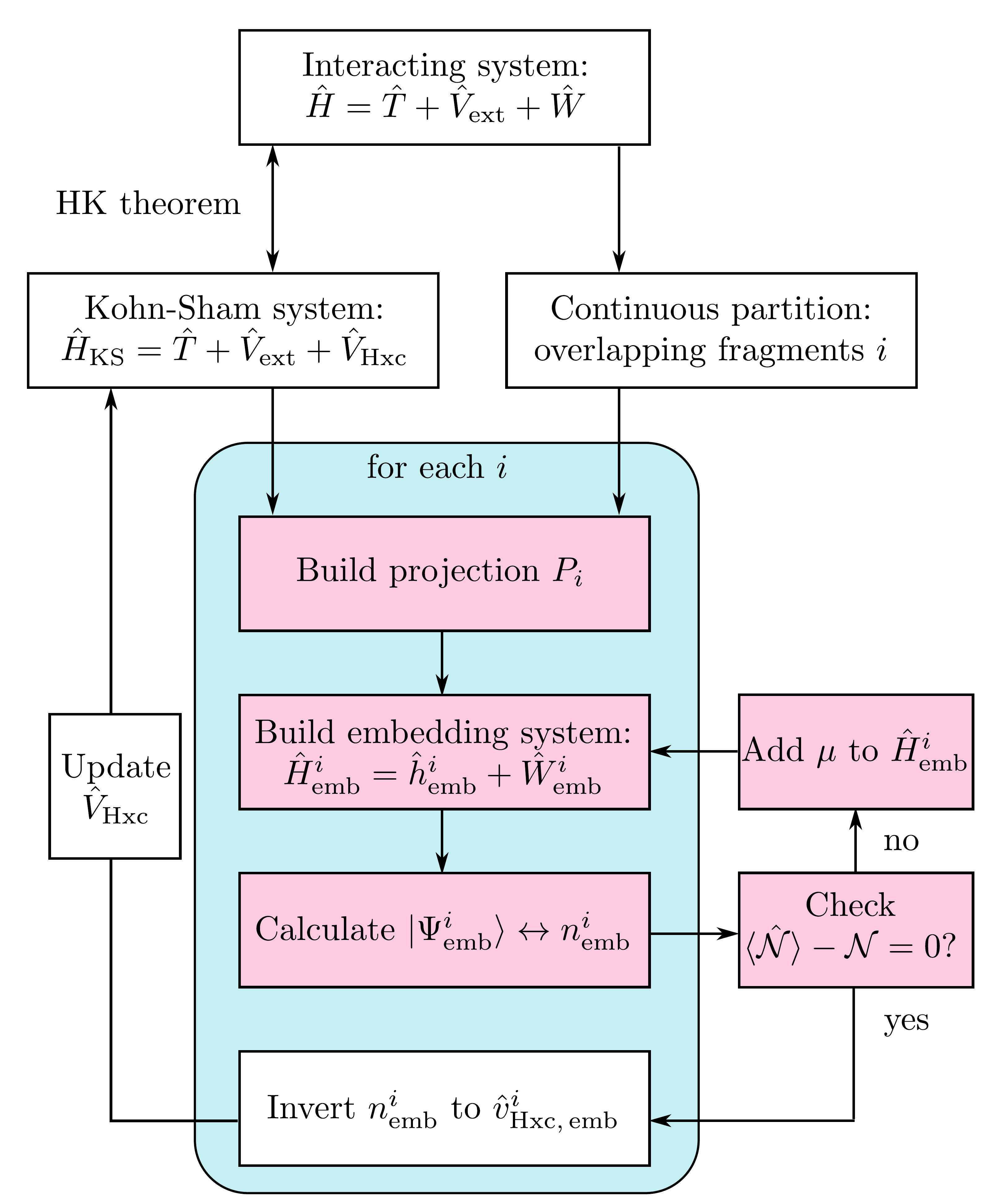}
	\caption{Visualization of the SDE algorithm: The full system can be uniquely mapped onto a non-interacting KS system. The system is divided into overlapping fragments such that a continuous reconstruction of the full system is possible. An initial guess
        for the global KS system is made, from which a projection is build for each fragment. Then, for each fragment the embedding Hamiltonian is calculated and the corresponding ground-state wave function and density are computed. A self-consistency cycle is added to maintain the correct particle number. As soon as the correct particle number is ensured in the full system, the density of every fragment is inverted and yields an updated $v_{\rm Hxc}$ on each site independently. This potential is then used to update the KS system. The procedure is repeated until self-consistency. In pink we mark those parts of the algorithm that are close to the DMET approach.}
        \label{pic:algorithm_loop}
\end{figure}

This yields the new KS Hamiltonian $\hat{H}_{\rm KS} = \hat{T}+\hat{V}_\mathrm{KS}$, which is then used to calculate a new set of projections $P_i$. This is done until
convergence (see algorithm in Figure \ref{pic:algorithm_loop}). 
Eventually, we obtain an accurate density and KS potential from which also correlated observables can be calculated as described in eq~(\ref{eq:observable}). The SDE algorithm can be improved systematically by increasing the fragment size and it converges to the exact solution. Note that the choice of reproducing accurately the density of the interacting embedded system by a non-interacting one is crucial as it is based on rigorous one-to-one relations between densities and potentials in DFT and gives us a well defined target for the inversion. This would not be the case with any other quantity such as e.g. the 1RDM (which is used in DMET), since the 1RDM of an interacting system cannot be reproduced exactly by a non-interacting one. 

As in SDE we adopt the projection from DMET, to make the distinction between the two more clear, we mark in Fig.~\ref{pic:algorithm_loop} in pink, which parts of the algorithm SDE shares with DMET. Both methods coincide only for fragment size $N_{\rm frag}=1$, as only then there is no difference in partition (single-site fragments cannot overlap) and also between density and 1RDM on the fragment (as there are no off-diagonal elements).

To complete the introduction of the SDE method, we now turn to its numerical cost. The cost of fragment calculations in SDE grows
exponentially with the fragment size $N_{\rm frag}$ and the cost for the underlying calculation of the non-interacting system grows quadratically
with the total number of grid points $N$. This has to be multiplied by the number of fragments, which is also $N$, and the needed self-consistency iterations $\eta$ yielding a total scaling of $4^{2 \cdot N_{\rm frag}}\cdot N^3\cdot \eta $. This is, of course, more expensive than a usual DFT calculation (that is $N^2\cdot \eta$ in local density approximation (LDA)), but cheaper than the exponentially growing cost of a FCI calculation.

\section{Diatomic molecule model in one and two dimensions}\label{section:H2}
In this section we introduce the model Hamiltonians, which we use to validate our approach (see section~\ref{sec:results}), and also the inversion scheme used for all results. 

\begin{figure}[t]
 \includegraphics[width=.4\textwidth]{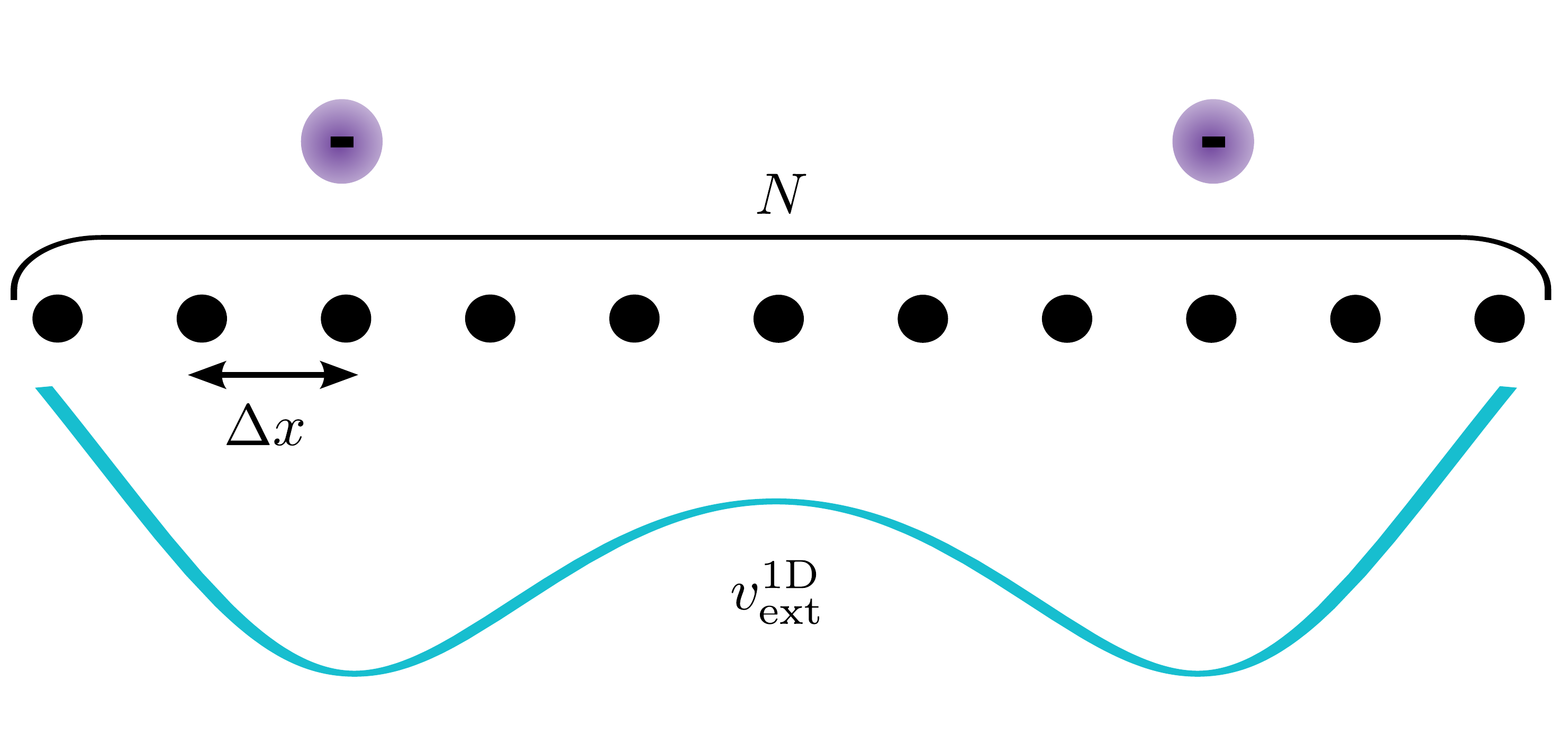}
        \caption{Visualization of the 1D $H_2$ molecule. The real space is
        discretized on a grid with $N$ sites. The two atoms are modeled through a symmetric
        double well potential $v^\mathrm{1D}_\mathrm{ext}$.}
 	\label{pic:H2}
 \end{figure}

The SDE approach so far is valid for all closed systems that can be represented by a time-independent Schr\"odinger equation. In order to benchmark our method and to show its efficiency, we describe the two-electron bond stretching of a heteroatomic
molecule in one and two dimensions (see Fig. \ref{pic:H2}).

We model this system with the following
Hamiltonian \footnote{We use atomic units (a.u.) throughout the paper.} on a 1D/2D real-space grid \cite{lubasch_systematic_2016}
 \begin{align}\label{eq:ham}
 \hat{H} & = -\frac{1}{2 \, \Delta x^2} \sum_{i, \sigma}  (\hat{c}^\dagger_{i+1,\sigma}\hat{c}_{i,\sigma} + \hat{c}^\dagger_{i,\sigma}\hat{c}_{i+1,\sigma} - 2 \hat{n}_{i,\sigma}) \nonumber\\
 & + \sum_{i, \sigma} v_{i,{\rm ext}} \hat{n}_{i, \sigma} 
 +  \sum_{i,j,\sigma,\sigma'}   \frac{\hat{n}_{i \sigma} \hat{n}_{j \sigma'}}{2\sqrt{\left(\Delta x (i-j)\right)^2+\alpha}},
 \end{align}
where $\hat{c}^\dagger_{i,\sigma}$ and $\hat{c}_{i,\sigma}$ are the usual creation and
annihilation operators of an electron with spin $\sigma$ on site $i$
and $\hat{n}_{i,\sigma} = \hat{c}^\dagger_{i,\sigma} \hat{c}_{i,\sigma}$ is the corresponding
density operator. In 2D the index $i$ becomes a double index with
 \begin{align}
  i & \rightarrow (i_x,i_y) \nonumber \\
  i+1 & \rightarrow (i_x+1,i_y),(i_x,i_y+1) \nonumber \\
  i-j & \rightarrow (i_x-j_x)^2 + (i_y-j_y)^2 
 \end{align}
The spacing $\Delta x$ is determined by the box size $L$ in direction $x$ 
and the number of grid points $N$ and as external potential we employ a
double well potential $v_{\rm ext}$.

The first part of the Hamiltonian takes
into account the kinetic energy of the molecule by means of a next-neighbors
hopping term.  The second term in eq~(\ref{eq:ham}) is the external potential which mimics the ions
of the molecule and depends on the considered dimension.  In the one
dimensional case, the external potential on each point is given by
\begin{equation}
	v^{\rm 1D}_{i,\rm ext} = -\cfrac{z_1}{\sqrt{\left(x_i - \frac{d}{2}\right)^2+\alpha}} -\cfrac{z_2}{\sqrt{\left(x_i + \frac{d}{2}\right)^2+\alpha}} + \cfrac{z_1 z_2}{2 \sqrt{\left(d^2 + \alpha \right)}}
\end{equation}
with $x_i = \Delta x \left(i-\frac{N-1}{2}\right)$. The numbers $z_1$ and $z_2$
determine the depth of each well respectively. In our case they take values
between $0$ and $2$ and we will characterize the potential by their difference
$\Delta z = z_1 - z_2$.  In the two-dimensional case the external potential
takes the form 
\begin{equation}
	v^{\rm 2D}_{i,\rm ext} = v^{\rm 1D}_{i_x,\rm ext} \cdot \cfrac{1}{\sqrt{\Delta x^2\left(i_y-\frac{N_y-1}{2}\right)^2 +\alpha}}
    \label{eq:vext_2d}
\end{equation}
accounting for both, the charge distribution of the ions in $x$ and $y$ direction.

The third term of the Hamiltonian takes into account the interaction 
of the electrons.  We model the electronic interaction as well as the core
potentials by the soft-Coulomb interaction, which avoids the
singularity at zero distance. In order to do so, we include a softening parameter $\alpha =1$. 

One reason for choosing a problem that only includes two electrons is that for
this example we can analytically invert the density $n$ of the interacting problem
to yield the potential $v_{\rm S}[n]$ of the auxiliary non-interacting system that has the same density. As the ground state of a two-electron problem is always a singlet it is valid that 
 \begin{align}
 n(\boldsymbol{r}) = 2 \left|\varphi_0(\boldsymbol{r})\right|^2.
 \end{align}
Inserting this property into the one-body equations eq~(\ref{eq:ks_eq}) yields \cite{helbig_exact_2009}
 \begin{align}\label{formula:inv}
 	\hat{v}_{\rm Hxc}[n](\boldsymbol{r})= \frac{\left( \boldsymbol{\nabla}^2/2  - v[n](\boldsymbol{r})\right)\sqrt{n(\boldsymbol{r})}}{\sqrt{n(\boldsymbol{r})}}+\varepsilon_0.
 \end{align}
Where $v[n]$ is the external potential of the interacting system which yields the same density $n$. The constant $\varepsilon_0$ can be chosen arbitrary as it only fixes the gauge. We choose it such that $\hat{v}_{\rm Hxc}(\boldsymbol{r})$ vanishes at the boundaries. The formula above is given in the real space domain but it can be applied to any quantum lattice system\footnote{Of course, the formula has to be adapted to the Hamiltonian of this system.}, as there is a one-to-one correspondence between density and potential for those systems \cite{Chayes1985}.
The exact inversion formula can therefore be applied to every embedded system with two electrons, hence, to every embedded system resulting from our model.
Note that although the exact inversion formula can only be used for the special case of two electrons, there are different ways to expand this towards the treatment of more particles.
The analytic inversion can either be replaced by numerical inversion schemes \cite{kananenka_efficient_2013, Jensen2017, Nielsen2018} or by the robust minimization scheme used in DMET \cite{wouters_practical_2016}.

\section{Results}
 \label{sec:results}
 
 To demonstrate the feasibility of our approach we calculate densities, KS potentials and total energies of model Hamiltonians introduced in section~\ref{section:H2}. Although our numerical results are limited to 1D and 2D model systems, we still discuss cases that are notoriously difficult to capture for standard KS DFT.
\subsection{Dissociation of the one-dimensional $H_2$ molecule}

Common DFT functionals like the local density approximation (LDA \cite{kohn_self-consistent_1965}),
or generalized gradient approximations (GGA \cite{perdew_generalized_1996,becke_density-functional_1988})
fail to describe the dissociation limit of the $H_2$ molecule. This failure is attributed to the so-called static correlation error, which is related to fractional spin states \cite{cohen2008insights}.
 Common approximate functionals, however, violate this condition and predict wrong energies for fractional spin states resulting in the wrong dissociation limit.\par

Although there are methods such as the strictly-correlated electron functional \cite{vuckovic_hydrogen_2015}, functionals based on the random phase approximation (RPA) \cite{heselmann_correct_2011,Fuchs2005} and on $GW$ combined with RPA \cite{Hellgren2015},
or the exchange-correlation potential by Baerends \emph{et.~al.} \cite{vanLeeuwen1994,Gritsenko1995}, which were designed to overcome these issues,
modeling the bond stretching of $H_2$ remains a challenging test for any new functional.

\begin{figure}[t]
\includegraphics[width=.5\textwidth]{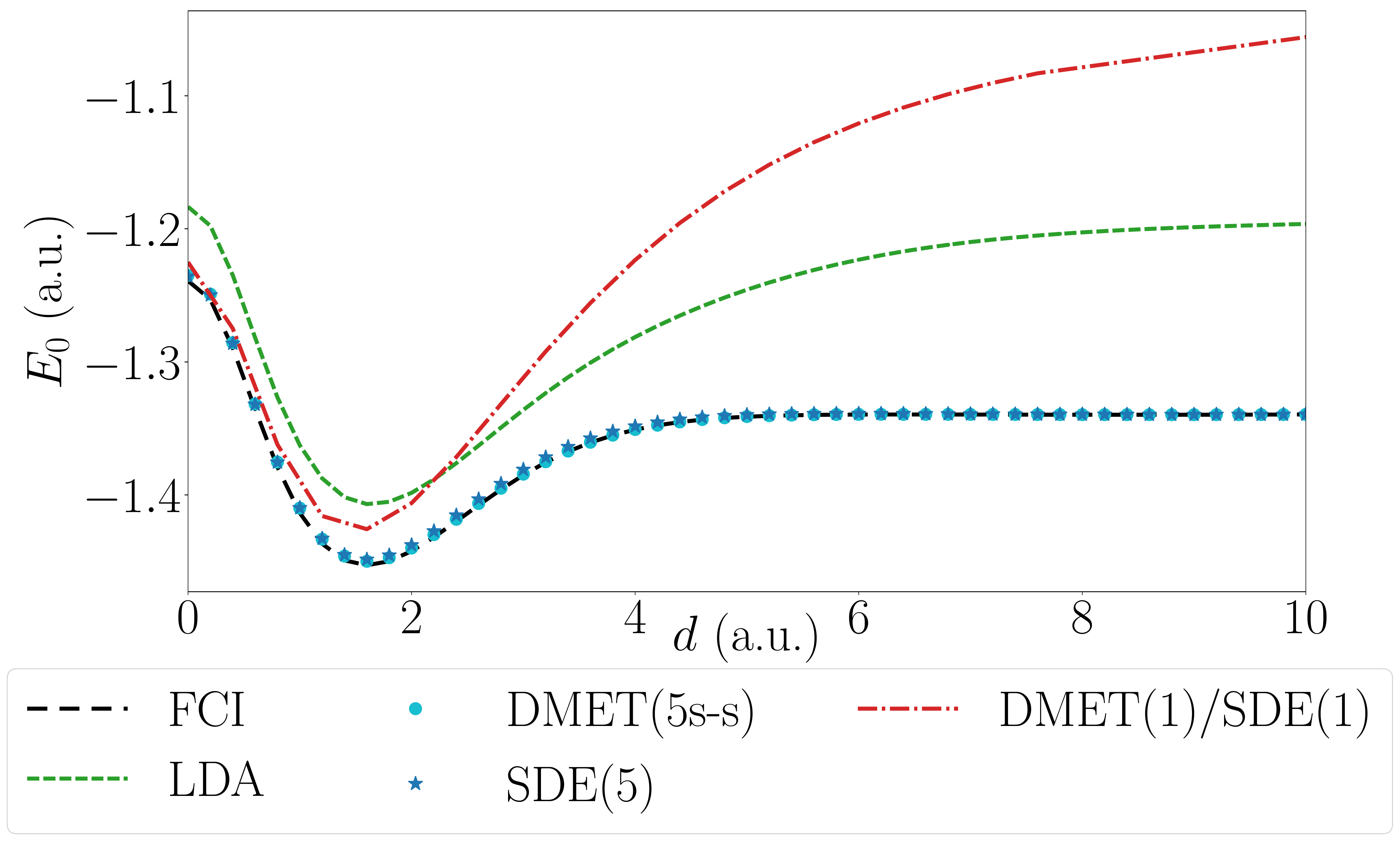} 
 	\caption{Ground-state energy of the 1D $H_2$ molecule, calculated with FCI (black dashes line), one dimensional LDA (green dashed line), five-sites single-shot DMET (turquoise circles), five-sites SDE (blue stars) and single-site DMET/SDE (red dash-dotted line). While LDA and DMET(1)/SDE(1) fail to describe the correct long-distance behavior, both DMET(5s-s) and SDE(5) show excellent agreement with the exact result. The following set of parameters has been used (see section \ref{section:H2}): number of real space grid sites $N=120$, box size $L=20$, potential well difference $\Delta z =0$, softening parameter $\alpha = 1$.
 	Atomic units (a.u.) are used throughout the paper.}
\label{graph:LDA}
\end{figure}

In Fig.~\ref{graph:LDA}, we show how the SDE method performs in this test case. We plot the ground-state energy of the Hamiltonian in  eq~(\ref{eq:ham}) with $\Delta z = 0 $ as function of interatomic distance calculated with FCI, one dimensional LDA-DFT \cite{helbig_density_2011}, one-site DMET (DMET(1)) that is equivalent to one-site SDE (SDE(1))\footnote{DMET(1) is the only version of DMET that we could apply to the model systems studied here self-consistently. It is also the only case in which DMET and SDE results coincide (for details, see section~\ref{sec:scf}).},  single-shot DMET with five fragment sites (DMET(5s-s)),
and five-site SDE (SDE(5)). The initial guess for the projection for both SDE and DMET is build from the one-body part of the Hamiltonian in eq~(\ref{eq:ham}).
The exact (FCI) energy curve shows the following well-known behavior: when varying the distance of the two core potentials $d$, the curve has a minimum corresponding to a stable molecule. For smaller core distances, the energy grows due to the repulsion of the two cores. Increasing the distance $d\rightarrow \infty$ leads to the vanishing of the binding energy resulting in two separate atoms.
 	
As discussed above, LDA does not predict the correct dissociation behavior of $H_2$ due to the static correlation error, the energy of the two separated atoms is overestimated. One-site embedding methods DMET(1)/SDE(1) also fail to describe this behavior correctly as static correlation cannot be captured with such small fragment sizes. They perform even worse than LDA for large distances.

In contrast, both SDE and single-shot DMET show excellent agreement with FCI for $N_{\rm frag}=5$. Both curves are on top of the FCI result. DMET even results in slightly better energies for intermediate distances. This might seem surprising at first glance, but the SDE algorithm is optimized to provide good densities and potentials and,
as widely discussed in the literature \cite{Medvedev49}, this does not necessarily go hand in hand with more accurate energies. The difference in energy between SDE and DMET is, however, negligible and in the next section we show that SDE, indeed does provide excellent densities and KS potentials. 
\color{black}

\subsection{Peaks and steps in the KS potential}\label{sec:KS}
\begin{figure}[t]
 \includegraphics[width=.5\textwidth]{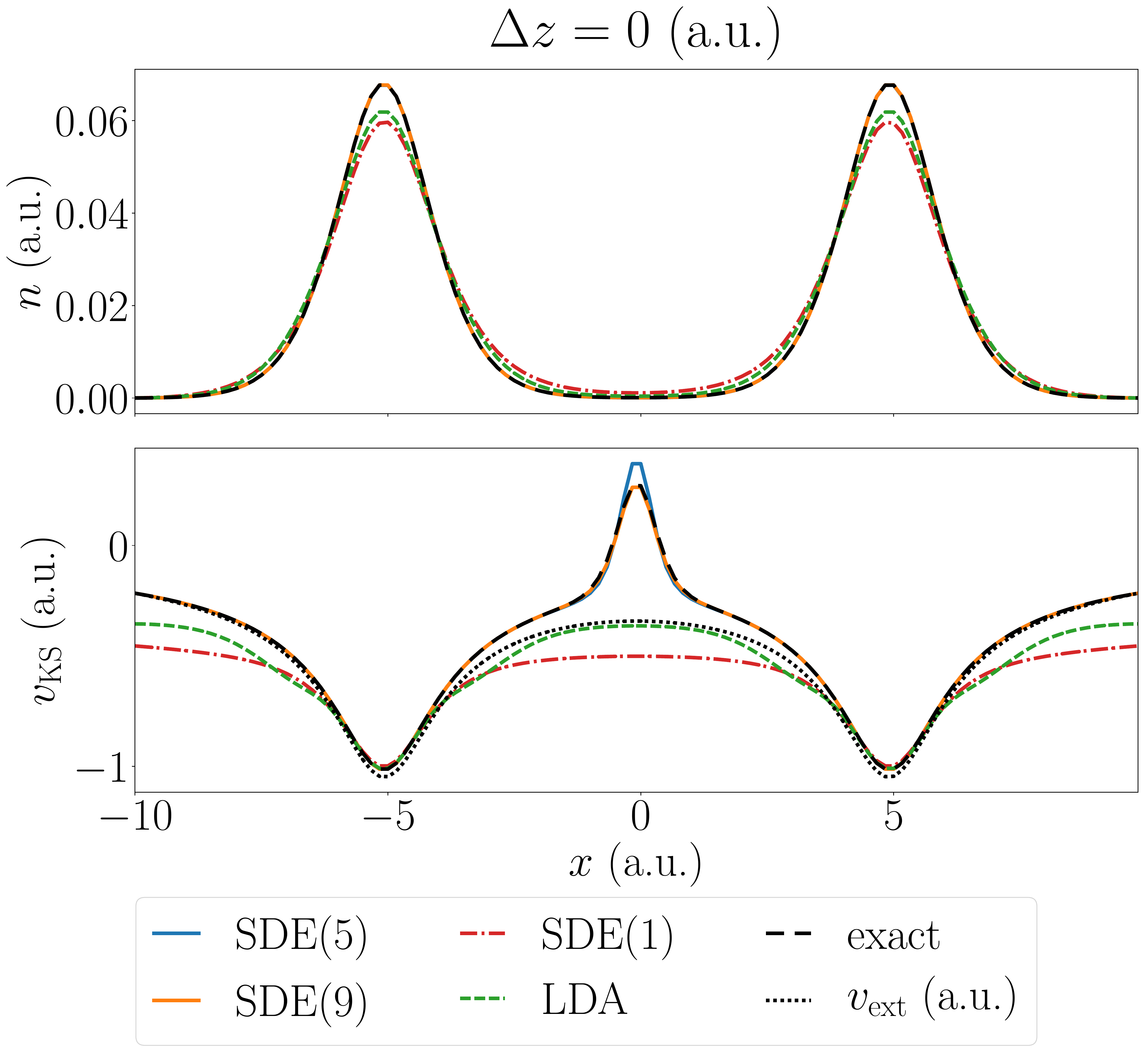}
        \caption{Density distribution $n(x)$ and KS potential $v_\mathrm{KS}(x)$
        with SDE(5) (blue solid line), SDE(9) (orange solid line), SDE(1) (red dash-dotted line), LDA (green dashed line) and FCI (black dashes line). The exact and the SDE solutions for fragments sizes larger than one agree quantitatively. The SDE KS potential in these cases shows the expected peak in the center which mimics the electron-electron interaction. For $N_{\rm frag} = 5$, this peak is slightly  overestimated, but converges quickly to a quantitatively exact
        result for $N_{\rm frag} = 9$. The SDE(1) and LDA results on the other hand differ significantly from the exact solution. The peak in the KS potential is missing completely. The following set of parameters has been used: $N=120$, $L=20$, $d = 10$.}
 	\label{graph:Peaks}
 	
 \includegraphics[width=.5\textwidth]{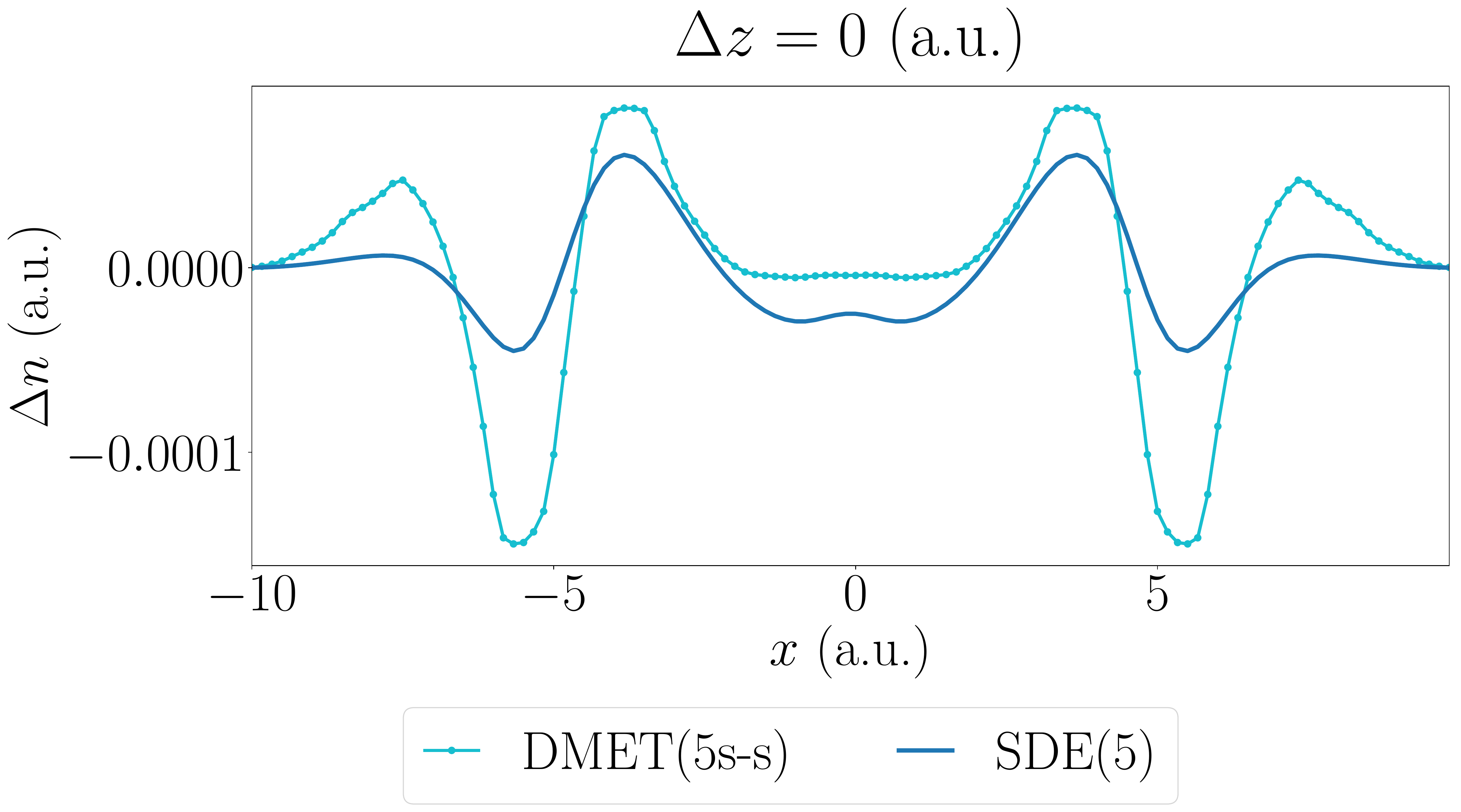}
        \caption{Deviation of densities $\Delta n$ from FCI reference results for five-site SDE (blue solid line) and five-site single-shot DMET (turquoise solid line with circles). SDE density exhibits smooth behaviour while DMET density shows discontinuities at fragment boundaries. The following set of parameters has been used: $N=120$, $L=20$, $d = 10$.}
 	\label{graph:diff2dmet}
\end{figure}

For the $H_2$ model, the KS system needs to describe the repulsion of the
two electrons. As the system does not include an actual
interaction term, this repulsion needs to be mimicked by the KS
potential.  As has been investigated in various works \cite{vanLeeuwen1994,Gritsenko1995,helbig_exact_2009}, 
we expect to see a peak that prevents the two electrons
from being at the same atom.
In Fig.~\ref{graph:Peaks}, we plot the density and the KS potential
obtained with SDE for fragment sizes of 1,  5 and 9 sites and compare them with the exact
density and the exact KS potential. 

\renewcommand{\textfraction}{0.05}

The density from the SDE calculations for the two larger fragment sizes agrees
quantitatively with the exact density. We also see a peak at position $x=0$ in
the KS potential for both SDE(5) and SDE(9) calculations. This peak is slightly overestimated for $N_{\rm frag}=5$, but agrees quantitatively with the exact solution as the fragments gets bigger ($N_{\rm frag}=9$). The SDE(1)/DMET(1) results are also plotted. As already discussed in the case of the energy, both density and potential deviate strongly from the exact solution. The peak in the KS potential accounting for strong correlations in the system is missing completely and, hence, also the  density distribution deviates strongly from the exact solution. The same applies to results obtained with LDA.\par

Further, we compare SDE densities to the ones from our real-space implementation of single-shot DMET that showed good results for ground-state energies of the model in the previous section. In Fig.~\ref{graph:diff2dmet}, we plot the deviation of the approximate densities $\Delta n$ from the exact ones (FCI) for both methods for $N_{\rm frag}=5$. We see that the DMET density deviates stronger from the exact solution than the SDE density. Furthermore, in DMET we clearly see a peculiarly shaped density, especially at fragment boundaries. This behavior is caused by the fact that there is no smooth connection between the fragments. This comparison reveals the need of our type of partitioning in order to have accurate densities.

\begin{figure}[t]
 \includegraphics[width=.5\textwidth]{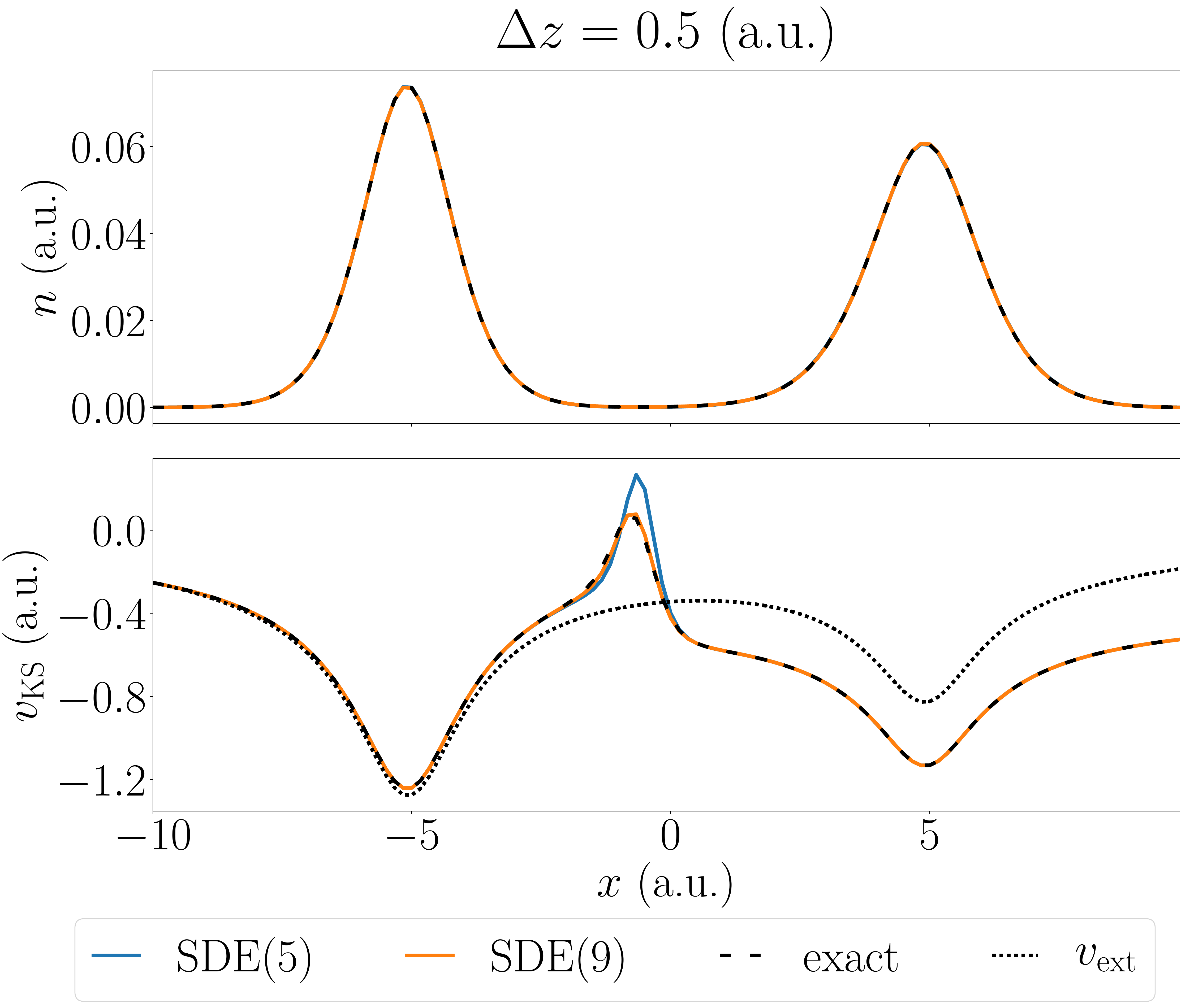}
        \caption{Density distribution $n(x)$ and KS potential $v_\mathrm{KS}(x)$
        for an asymmetric external potential with SDE(5) (blue solid line), SDE(9) (orange solid line) and FCI (black dashes line). Both SDE results agree with the exact solution and show expected peak and step in the KS potential. The following set of parameters has been used: $N=120$, $L=20$, $d = 10$.}
 	\label{graph:Steps}
\end{figure}

As the next challenge we consider more general situations such as bond stretching of heteroatomic molecules, such as $LiH$, that can also be modeled by the Hamiltonian of eq~(\ref{eq:ham}) by considering an asymmetric external potential. The SDE results are plotted in Fig.~\ref{graph:Steps} and also here we observe excellent agreement with exact results for both density and potential. We observe an asymmetric density distribution, which is mimicked by a KS potential that, in addition to the peak observed in the symmetric case in Fig.~\ref{graph:Peaks}, has a step between the two wells. The appearance of the step and its importance in KS DFT is to this day a widely discussed issue in the literature\cite{elliott2012universal,luo2013absence,hodgson2016origin, hodgson2017interatomic}.

Even though approximate functionals, e.g. those based on the exact-exchange approximation, do reproduce the step in the KS potential\cite{schonhammer1987discontinuity}, to the best of our knowledge, so far there does not exist any approximate energy functional that can reproduce both peaks and steps\cite{Hellgren2019} at the same time. Within the 
SDE approach we achieve both claims and that is why we believe that with SDE we provide a new path towards accurate KS potentials even for strongly correlated systems.

 \subsection{Convergence behavior}

\begin{figure}[t]
	\includegraphics[width=.5\textwidth]{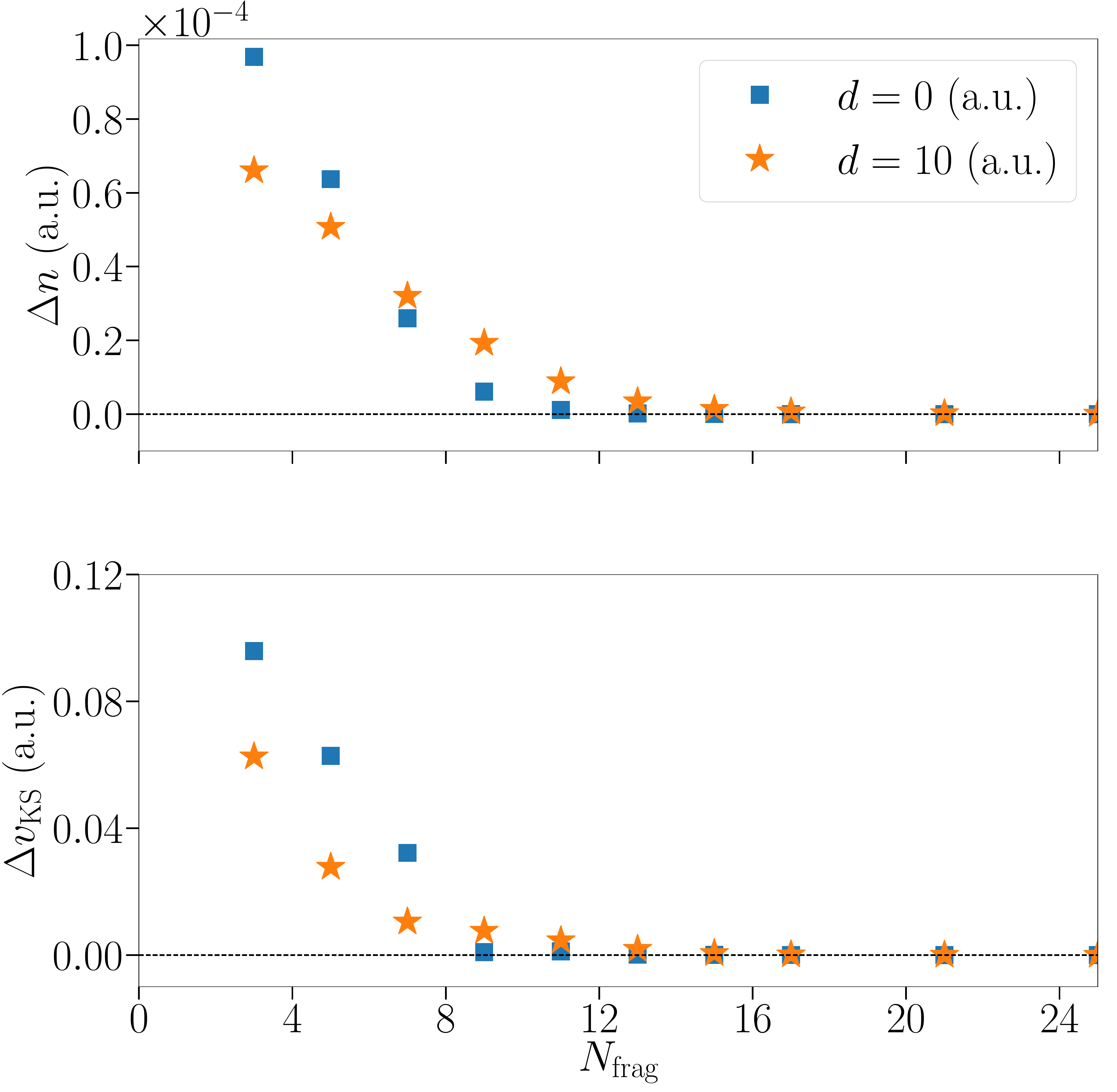}
        \caption{Integrated deviation of the density (upper graph) and the
        KS potential (lower graph) of the SDE calculation from the exact
        solution for weakly static correlated ($d=0$) and strongly static correlated electrons
        ($d=10$). In both cases, we observe a decrease in the error between the
        two calculations. While in the weakly correlated case the error
        estimate is higher for small fragments and decreases faster, in the
        strongly correlated case already the calculations for small fragments
        are very good and decrease slower. Already for ($N_{\rm frag} = 3$), the
        error is of the order of $\Delta n \le 10^{-4}$.  Parameters for $d=0$:
        $N=120$, $L=10$, $\Delta z = 0$, $\alpha = 1$; parameters for $d=10$:
        $N=120$, $L=20$, $\Delta z = 0$, $\alpha = 1$}
 	\label{graph:convergence_dens}
 \end{figure}

\begin{figure}[t]
 \includegraphics[width=.5\textwidth]{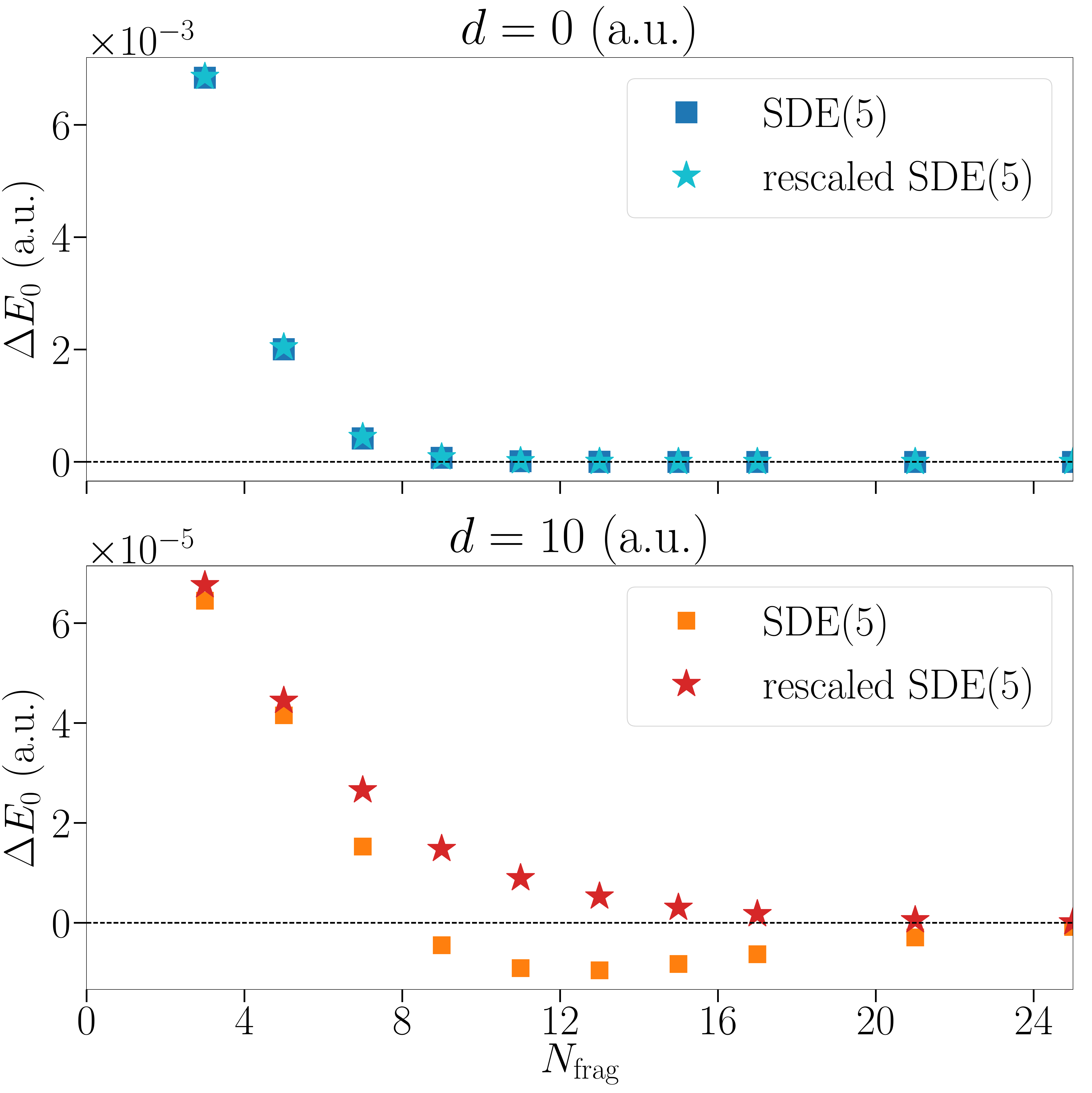}
        \caption{Difference of the total energy between the SDE and the exact
        solution $\Delta E_0$ with and without rescaling with respect to the
        particle number. We consider two different core distances ($d=0$, upper
        graph and $d=10$, lower graph), which correspond to weak and strong
        correlation between the electrons.  For the weakly static correlated system,
        already for $N_{\rm frag} = 9$, the error between the two calculations
        is below our selected accuracy limit. For strongly static correlated electrons, $d=10$,
        we observe that the energy estimate of the SDE calculations for $N_{\rm
        frag}\ge 9$ is too low compared to the exact solution. The deviation in
        energy is very low for small fragment sizes $(\Delta E_0 \le 10^{-5})$.
        Parameters for $d=0$: $N=120$, $L=10$, $\Delta z = 0$, $\alpha = 1$;
        parameters for $d=10$: $N=120$, $L=20$, $\Delta z = 0$, $\alpha = 1$}
 	\label{graph:convergence_eng}
 \end{figure}

In contrast to conventional DFT approaches, the SDE method can be improved 
systematically simply by increasing the size of the fragments. In Fig.~\ref{graph:convergence_dens} and
\ref{graph:convergence_eng}, we see the deviation of our results from the exact
solution for different properties $Q$ of the system, integrated over the whole
space:
 \begin{align}
 	\Delta Q= \sum_i \left| Q_i^{\rm SDE} - Q_i^{\rm exact} \right|\cdot \Delta x,
 \end{align}
 where $\Delta x$ is the grid constant.
 
 In Fig.~\ref{graph:convergence_dens}, we plot the deviation of the density $\Delta n$ and the KS potential $\Delta v_{\rm KS}$ between the SDE calculation and the exact result. 
We consider two different core distances ($d=0$ and $d=10$), which correspond
to weak and strong static correlation between the electrons. In both cases and for
both chosen properties, we observe a monotonous decrease in $\Delta Q$ with increasing
fragment size up to a quantitative agreement of the two solutions. Already for
the smallest considered fragment size $N_{\rm frag} =3$, the deviations are
relatively small, that is of the order of the fourth digit for the density
$\Delta n \le 10^{-4}$ and of the order of the first digit for the KS
potential $\Delta v_{\rm KS}=10^{-1}$.

In Fig.~\ref{graph:convergence_dens},
we show the deviation of the total energy $E_0$ of the SDE method from the
exact calculation.
Again, we consider one example with weakly static correlated electrons and one example
with strongly static correlated electrons. For weakly correlated electrons, the
difference in energy decreases and already for an fragment size of $N_{\rm frag}
= 7$, the deviation from the exact solution is below chemical accuracy of $1.6 \ \mathrm{m hartree}$.

For strongly (static) correlated electrons, we observe that the SDE energy becomes
smaller than the exact energy for a range of fragments between $N_{\rm frag} =
9$ and $N_{\rm frag} =20$. This is because the SDE method is not variational and the estimate for the energy therefore can also be lower than the exact energy. Also for this observable though, already for small fragments our estimate is of order $\Delta E_0 \le 10^{-5}$ which is far below chemical accuracy.

Since we approximate the wave function of the full system by a set of fragment wave functions, the total particle number calculated with fragment wave functions is not necessarily correct. The employed optimization of the chemical potential leads to the correct number for $\langle\hat{\mathcal{N}}\rangle$ up to a desired accuracy ($|\langle\hat{\mathcal{N}}\rangle - \hat{\mathcal{N}}| < 10^{-5}$).
As the energy difference is of the same order of magnitude, we further rescale the energy with respect to the particle number
\begin{align}
E^{\rm SDE}_0 \rightarrow E^{\rm SDE}_0 \cdot \hat{\mathcal{N}}/\langle\hat{\mathcal{N}}\rangle,
\end{align}
to see if we achieve a better convergence behavior. We indeed do, as we can also see in Fig.~\ref{graph:convergence_eng}. Nonetheless, the calculated energy can still be lower than the exact energy, meaning that we still observe the non-variational nature of our approximation.

\subsection{Application to systems in 2D}
In order to demonstrate that the SDE method can be applied to higher-dimensional models, we here discuss the
$H_2$ molecule and a model heteroatomic molecule in two dimensions.

\begin{figure}[t]
 \includegraphics[width=.5\textwidth]{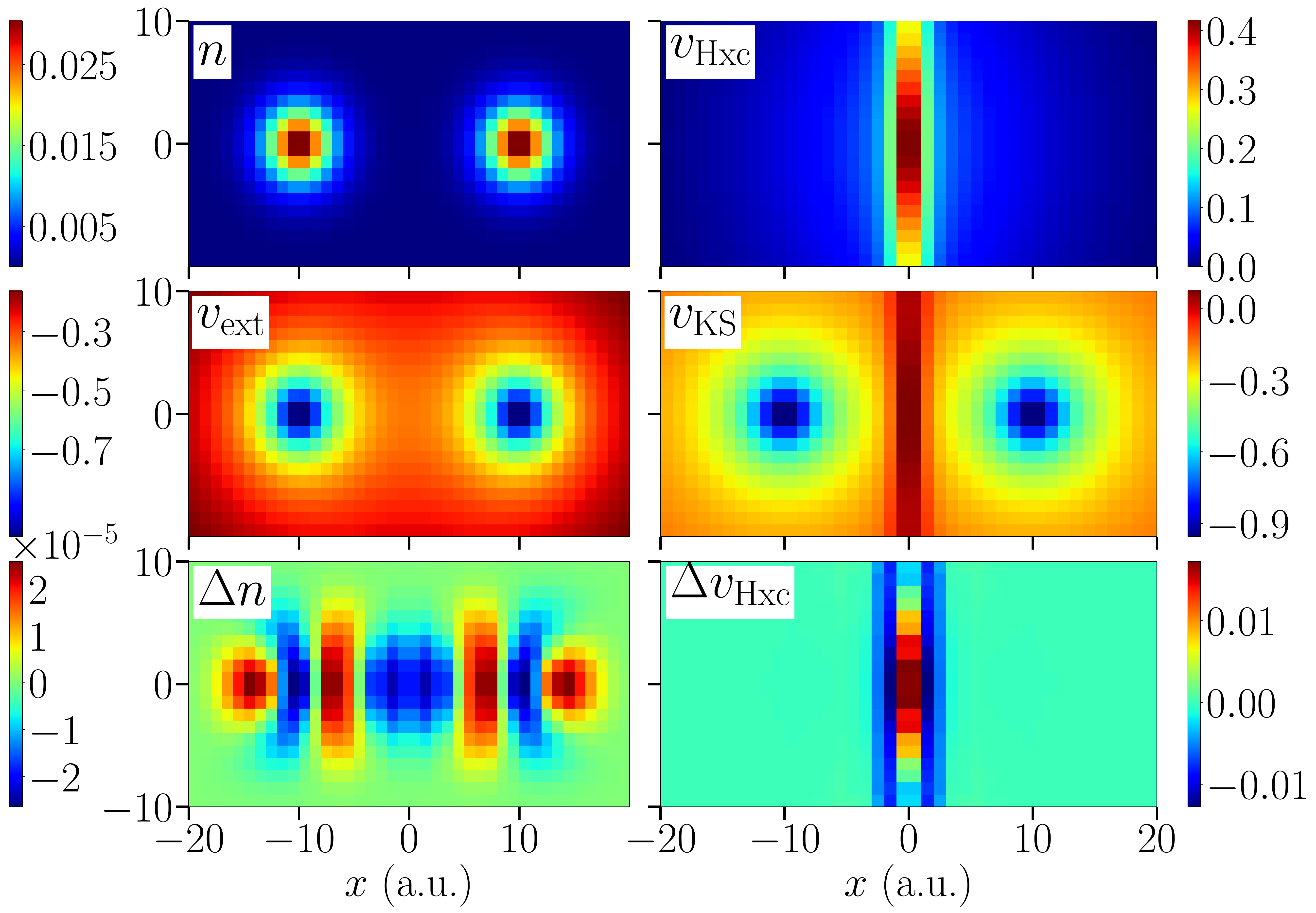}
        \caption{The $H_2$ molecule in two dimensions. Plotted are the
        density $n$, the Hartree-exchange-correlation potential
        $v_{\rm Hxc}$, as well as their difference from the exact reference $\Delta n$ and $\Delta v_{\rm Hxc}$, respectively, the KS potential $v_{\rm KS}$, and the external
        potential $v_{\rm ext}$ with SDE($4\times 4$). We observe a homogeneous density
        consistent with the external potential. $v_{\rm Hxc}$ shows the peak
        accounting for the interactions of the two electrons. We observe good agreement with the exact reference. The following set of
        parameters has been used: $N_x=40$, $N_y=20$, $L_x =20$, $L_y =10$, $d
        = 10$, $\Delta z = 0$} \label{graph:twod_peaks}
\end{figure}

In Fig.~\ref{graph:twod_peaks}, we plot the
density $n$, the KS potential $v_{\rm KS}$, the external potential
$v_{\rm ext}$, the Hartree-exchange-correlation potential $v_{\rm Hxc}$, and deviations from the exact solution $\Delta n$ and $\Delta v_{\rm Hxc}$ for
the two-dimensional $H_2$ model.
We observe a homogeneous density distribution around the two core potentials
that is consistent with the external potential. The Hartree-exchange-correlation potential which mimics the interactions of the electrons as well the kinetic correlations in the
interacting case, shows a peak in the middle of the molecule. Our observations
are consistent with the exact solution of this problem.

For a model heteroatomic molecule, we plot the same properties as for $H_2$.
\begin{figure}[t]
 \includegraphics[width=.5\textwidth]{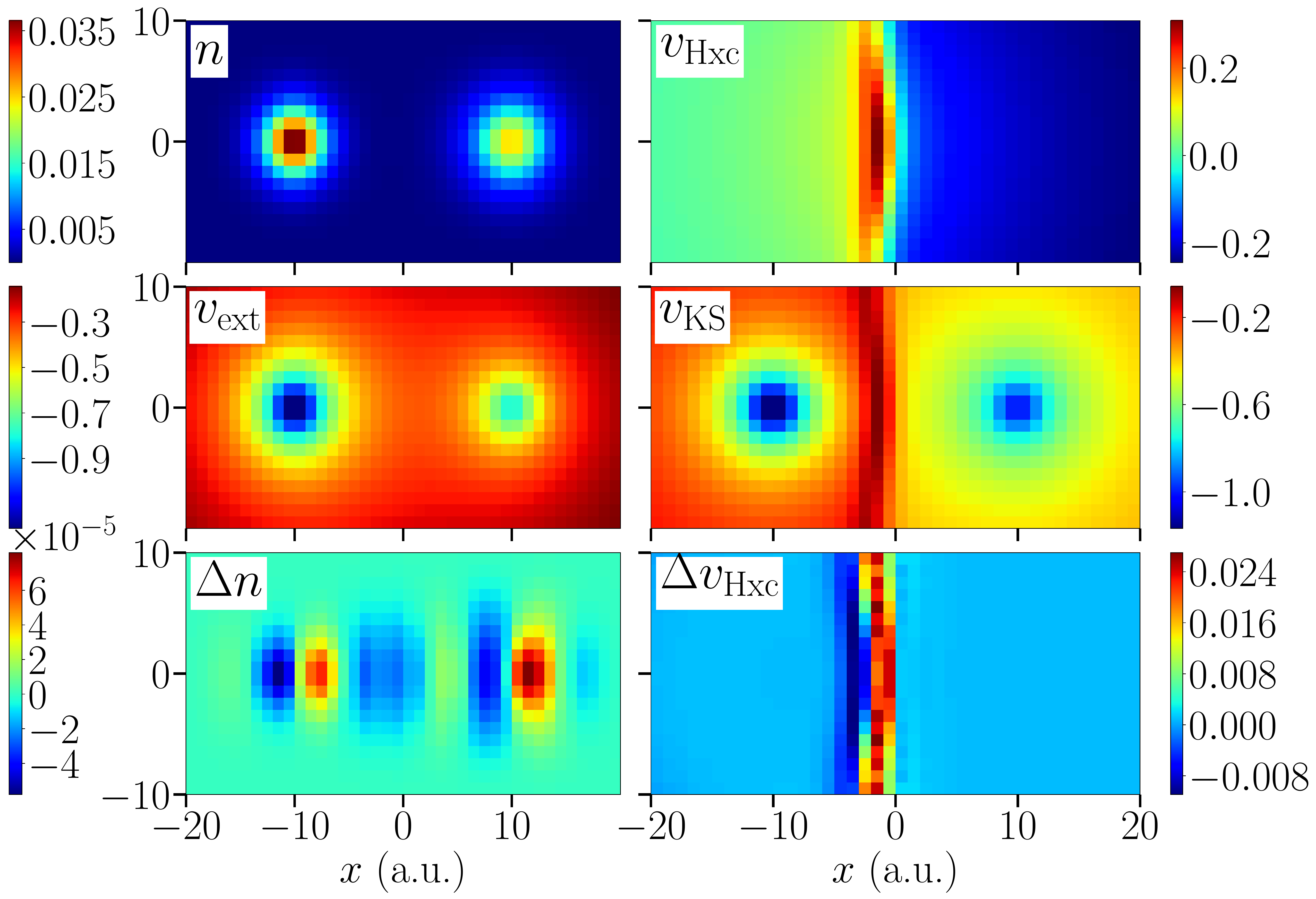}
        \caption{Heteroatomic molecule in two dimensions. Plotted are the
        density $n$, the Hartree-exchange-correlation potential
        $v_{\rm Hxc}$, as well as their difference from the exact reference $\Delta n$ and $\Delta v_{\rm Hxc}$, respectively, the KS potential $v_{\rm KS}$, and the external
        potential $v_{\rm ext}$ with SDE($4\times 4$). We observe an asymmetric density
        consistent with the external potential. $v_{\rm Hxc}$ again shows the
        peak accounting for the interactions of the two electrons.
        Additionally, a step accounting for the asymmetric distribution of the
        density can be observed. Again, we observe good agreement with the exact reference. The following set of parameters has been used:
        $N_x=40$, $N_y=20$, $L_x =20$, $L_y =10$, $d = 10$, $\Delta z = 0.5$}
 	\label{graph:twod_steps}
\end{figure}
The density for the heteroatomic molecule in the two-dimensional case is
asymmetrically distributed between the two cores, again consistent with the
external potential. In the Hartree-exchange-correlation potential, additional
to the peak accounting for the interaction of the electrons, we also observe a
step that accounts for the asymmetric distribution of the density.

\section{Conclusion and Outlook}
\label{sec:sum}

We present a self-consistent density-functional embedding (SDE) approach, which is a way to apply KS DFT without any explicit functional expressions but approximating the density to potential mapping. Observables $O$ are calculated through a set of fragment wave function which avoids the need of explicit functionals $O[n]$. SDE
yields accurate results for two-electron systems in one- and two dimensions
for moderate fragment sizes. Not only we can very accurately reproduce the exact potential energy surfaces of these systems, but also the peaks and steps in the KS
potential predicted by the exact solution. Additionally, the SDE method is
systematically improvable by increasing the size of the fragment and converges
to the exact solution.

To calculate larger fragment sizes and particle numbers with SDE a wide range of solvers based on DMRG \cite{schollwoeck_density-matrix_2005,hubig2017symmetry,chan2004algorithm},
coupled cluster \cite{bartlett2007coupled,sun2018pyscf,werner2012molpro}, selective CI \cite{sharma2017semistochastic}, or quantum Monte-Carlo \cite{booth2009fermion,haule2007quantum}
can be included into the algorithm. Further, in order to treat larger particle numbers,
the analytic inversion scheme in eq~(\ref{formula:inv}) has to be substituted 
by a numeric one, as e.g. proposed in \cite{kananenka_efficient_2013, Jensen2017, Nielsen2018}, or simply be replaced by robust optimization schemes as in conventional DMET \cite{wouters_practical_2016}. We expect to face one challenge with respect to the treatment of larger systems and that is the storage and projection of the electron-electron interaction term, which numerically is stored in a large tensor of fourth order (and thus also grows by fourth order with respect to the system size).
In order to treat larger systems, we either have to find an efficient way of storing the interaction 
tensor of the original system and then project it to the embedded system
or we could employ the non-interacting bath picture from DMET \cite{wouters_practical_2016}, that circumvents the 
treatment of the interaction tensor for the full system altogether.

In this work we provide a promising group of methods that combine functional methods with embedding schemes, yielding systematically improvable results. Work to extend the method to larger systems is underway.
\section{Acknowledgments}
\label{sec:appendix}
The authors acknowledge insightful discussions with M.~Ruggenthaler and C.~Sch\"afer.
U. M. acknowledges funding by the IMPRS-UFAST.
A.  R.  acknowledges financial support by the European Research Council  (ERC-2015-AdG-694097).  The Flatiron Institute is a division of the Simons Foundation.

\section{Appendix: The construction of the projection}
Here, we give step-by-step instructions on how the projection in DMET is constructed and how it is modified in SDE to account for different particle numbers in the system.

As discussed in section~\ref{sec:how_to}, the projection is nothing but a single-particle basis transformation optimized to describe the physics of the fragment.
The new basis is found as follows:

\begin{enumerate}
 \item The Hamiltonian of the full system with $M$ electrons is approximated by a non-interacting single-particle Hamiltonian $\hat{h}_{\rm mf}$\footnote{In DMET the choice of  $\hat{h}_{\rm mf}$ is not fixed and usually the Fock operator is used. In SDE, $\hat{h}_{\rm mf}$  is the single-particle operator of the KS system.}
 with corresponding single-particle eigenvalue equation $\hat{h}_{\rm mf}\varphi_j (\mathbf{r})= \varepsilon_j \varphi_j (\mathbf{r}) $. From this we calculate the
 the spin summed 1RDM in the grid basis. It is a $N \times N$ matrix that reads:
 \begin{align}
 \label{eq:mf1rdm}
 \gamma_{\mu\nu}=2\sum_{j=1}^{M/2} \varphi_j^{*}(\mathbf{r}_\nu) \varphi_j(\mathbf{r}_\mu).
 \end{align}
 The fact that only the lowest $M/2$ eigenvectors contribute is a direct consequence of the fact that this 1RDM is build from a non-interacting wavefunction. In case of an interacting one, all $N$ eigenvectors $\varphi_j (\mathbf{r})$ would contribute with some occupation number $\lambda_i$, which lies between $0$ and $2$. 
 \item \label{bullet:2} Having set up the 1RDM matrix $\gamma$, we separate it in different submatrix blocks, namely one that would correspond to those grid points that belong purely to the fragment, two blocks which contain one grid point on the fragment and one on the bath, and one block which has only bath grid points. 
  \begin{align*}
   \gamma & = \left(\begin{array}{@{}c|c@{}}
		    \begin{matrix}
		    \gamma_{1\,1} &  \cdots & \gamma_{1\,N_{\rm frag}} \\
		    \vdots 	  &  \ddots & \vdots	   \\
		    \gamma_{N_{\rm frag}\,1} &  \cdots & \gamma_{N_{\rm frag}\,N_{\rm frag}} \end{matrix} &
		    \begin{matrix}
		    \gamma_{1\, N_{\rm frag}+1} &  \cdots & \gamma_{1\,N} \\
		    \vdots 	  &  \ddots & \vdots	   \\
		    \gamma_{N_{\rm frag}+1\,1} &  \cdots & \gamma_{N_{\rm frag}+1\,N} \end{matrix} \\
		    \hline
		    \begin{matrix}
		    \gamma_{N_{\rm frag}+1\,1} &  \cdots & \gamma_{1\,N_{\rm frag}+1} \\
		    \vdots 	  &  \ddots & \vdots	   \\
		    \gamma_{N\,1} &  \cdots & \gamma_{N\,N_{\rm frag}+1} \end{matrix} &
		    \begin{matrix}
		    \gamma_{N_{\rm frag}+1\,N_{\rm frag}+1} &  \cdots & \gamma_{N_{\rm frag}+1\,N} \\
		    \vdots 	  &  \ddots & \vdots	   \\
		    \gamma_{N\,N_{\rm frag}+1} &  \cdots & \gamma_{N\,N} \end{matrix}
		  \end{array}\right) \\
     & \equiv \left(\begin{array}{@{}c|c@{}}
		    \gamma_{\rm frag} & \gamma_{\rm frag-bath} \\ \hline  \gamma^{\rm T}_{\rm frag-bath} & \gamma_{\rm bath}
		  \end{array}\right)
  \end{align*}
  Here, without loss of generality, we have assumed that the fragment contains the first $N_{\rm frag}$ grid points of the system.
 \item We then diagonalize the submatrix $\gamma_{\rm bath}$. Its eigenvalues $\widetilde{\lambda}_j$ will be all between zero and two containing up to $N_{\rm frag}$ eigenvalues with $0 < \widetilde{\lambda}_j <2$ \cite{macdonald1933successive}. 
 \item From the eigenvectors $\widetilde{\varphi}_{j, \, \mathrm{bath}}(\mathbf{r})$ with $0 < \widetilde{\lambda}_j <2$ we build the correlated bath orbitals of our CAS in real space basis as
 \begin{equation}
     \varphi_{j, \, \mathrm{corr.\, bath}}(\mathbf{r}_\mu) = \begin{cases}
                                                            0, & \text{if } \mu < N_{\rm frag},\\
                                                            \widetilde{\varphi}_{j, \, \mathrm{bath}}(\mathbf{r}_\mu), & \text{else.}
                                                           \end{cases}
 \end{equation}
 \item Having obtained in this way the correlated bath orbitals we also use a set of orbitals to describe the fragment. The fragment orbitals will be as many as the size of the fragment and each of them will have coefficient one at a specific fragment point and zero elsewhere.
 The embedding Hamiltonian $\hat{H}_{\rm emb}$ is constructed by projecting the full Hamiltonian $\hat{H}$ in the subspace that is spanned by the set of the aforementioned orbitals. In other words $\hat{H}_{\rm emb}$ is obtained by a basis transformation of the original Hamiltonian.

 \end{enumerate}

The number of correlated bath orbitals in the CAS is equal to $N_{\rm frag}$ as long as $2 N_{\rm frag} < M < 2\left(N- N_{\rm frag}\right)$ holds \cite{macdonald1933successive}, otherwise the number of correlated bath orbitals is smaller. As DMET was constructed for Hubbard-type lattice systems, for which the condition above mostly holds, in DMET the orbital construction that we just described is used without modifications. 

In SDE, we now modify the orbital construction of DMET in order to get $N_{\rm frag}$ correlated bath orbitals regardless of the particle number $M$. For the low particle numbers that are considered in this manuscript, we achieve this by artificially including correlations into the 1RDM of the full system by including higher-energy single-particle orbitals. In order to do so, we adjust the formula of eq~(\ref{eq:mf1rdm}) to
 \begin{equation}
 \label{eq:mf1rdm_mod}
     \gamma_{\mu \nu} = \sum\limits_{j=1}^{N_{\rm frag}} \varphi_j^{*}(\mathbf{r}_\nu) \varphi_j(\mathbf{r}_\mu) \cdot
     \begin{cases}
      2                         & \text{for } j < M/2 -1, \\
      2-\eta(N_{\rm frag} - M/2) & \text{for } j = M/2, \\
      \eta                      & \text{for } j > M/2,
     \end{cases}
 \end{equation}
 with some small value $\eta$ and then continue with the orbital construction from \ref{bullet:2}. The actual value of $\eta$ is not of great importance as it is only used to include higher-lying orbitals into the 1RDM and the same CAS would be obtained for different values of $\eta$. In our implementation $\eta=0.01$ is chosen.
 
 Note that eq~(\ref{eq:mf1rdm_mod}) is valid only for $M<2 N_{\rm frag}$. For large particle numbers $M>2\left(N- N_{\rm frag}\right)$ the procedure can be adapted in a straight-forward manner due to particle-hole symmetry.

%


\end{document}